\documentclass[twocolumn,secnumarabic,amssymb, nobibnotes, prapplied]{revtex4-2} 

\pdfoutput=1

\newcommand{\ket}[1]{\left|#1\right\rangle}
\usepackage{lipsum}
\usepackage{amsfonts,amssymb,amsmath}
\usepackage{graphicx}
\usepackage{bm}
\usepackage{xcolor,calc}
\usepackage{tikz}

\definecolor{color_comment}{rgb}{0.3, 0.8, 0.3}

\definecolor{color_out}{rgb}{0.7, 0.7, 0.7}

\definecolor{color_new}{rgb}{0.8, 0.3, 0.3}

\newcommand*\circled[1]{\tikz[baseline=(char.base)]{
    \node[shape=circle,draw,dashed,inner sep=1pt] (char) {#1};}}

\newcommand*\circleddd[1]{\tikz[baseline=(char.base)]{
    \node[shape=circle,draw,dashed,inner sep=3pt] (char) {#1};}}

\begin{document}

\title{Zero- and Low-Field Sensing with Nitrogen Vacancy Centers}
\author{Philipp J. Vetter${}^{1,2}$}
\email{philipp.vetter[at]uni-ulm.de}
\author{Alastair Marshall${}^{3}$}
\author{Genko T. Genov${}^{1,2}$}
\email{genko.genov[at]uni-ulm.de}
\author{Tim F. Weiss${}^{1,2}$}
\author{Nico Striegler${}^{3}$}
\author{Eva F. Großmann${}^{1,2}$}
\author{Santiago Oviedo-Casado${}^{4}$}
\author{Javier Cerrillo${}^{5}$}
\author{Javier Prior${}^{6}$}
\author{Philipp Neumann${}^{3}$}
\author{Fedor Jelezko${}^{1,2}$}
\affiliation{${}^{1}$Institute for Quantum Optics, Ulm University, Ulm D-89081, Germany
\\${}^{2}$Center for Integrated Quantum Science and Technology (IQST), Ulm D-89081, Germany
\\${}^{3}$NVision Imaging Technologies GmbH, Ulm D-89081, Germany
\\${}^{4}$Racah Institute of Physics, The Hebrew University of Jerusalem, Jerusalem 91904, Givat Ram, Israel
\\${}^{5}$\'{A}rea de F\'{i}sica Aplicada, Universidad Polit\'{e}cnica de Cartagena, Cartagena E-30202, Spain
\\${}^{6}$Departamento de F\'{i}sica, Universidad de Murcia, Murcia E-30071, Spain}

\begin{abstract}
Over the years, an enormous effort has been made to establish nitrogen vacancy (NV) centers in diamond as easily accessible and precise magnetic field sensors.
However, most of their sensing protocols rely on the application of bias magnetic fields, preventing their usage in zero- or low-field experiments.
We overcome this limitation by exploiting the full spin $S=1$ nature of the NV center, allowing us to detect nuclear spin signals at zero- and low-field with a linearly polarized microwave field.
As conventional dynamical decoupling protocols fail in this regime, we develop new robust pulse sequences and optimized pulse pairs, which allow us to sense temperature and weak AC magnetic fields and achieve an efficient decoupling from environmental noise.
%
Our work allows for much broader and simpler applications of NV centers as magnetic field sensors in the zero- and low-field regime and can be further extended to three-level systems in ions and atoms.
\end{abstract}

\maketitle


\section{Introduction}
Quantum sensing uses the quantum properties of systems such as solid-state defects, photons and ions, to estimate a physical quantity \cite{degen2017quantum}. 
The NV center in diamond is a successful example for such a 
quantum sensor \cite{balasubramanian2008nanoscale, mamin2013nanoscale, lovchinsky2016nuclear, schmitt2017submillihertz, aslam2017nanoscale}.
Most sensing protocols 
rely on bias magnetic fields that lift the 
degeneracy of their ground state manifold 
\cite{degen2017quantum}.
However, a bias magnetic field can lead to many undesirable effects, e.g., during structural analysis of molecules \cite{pelliccione2016scanned, thiel2016quantitative, jenkins2019single, glenn2018high}.
For example, it induces a Zeeman interaction which then dominates over the spin-spin coupling (J-coupling), masking crucial information about the chemical bonds.
%
%
%
%
Moreover, bias magnetic fields lead to perturbations in 
condensed matter systems, e.g., magnetic susceptibility effects \cite{joy1998relationship}. 
\newline
%
Recent work attempted to overcome this limitation by applying circularly polarized microwave fields 
to selectively address one spin state \cite{zheng2019zero, lenz2020magnetic}, while working at zero- or low-field.
However, this method requires special microwave structures to apply the circularly polarized microwave fields, whose performance strongly depends on their phase and placement 
relative to the NV center \cite{lenz2020magnetic, mrozek2015circularly}. 
\newline
We demonstrate a different approach, where we exploit the full spin $S=1$ nature of the NV center
for sensing at zero- and low-field.
By applying linearly polarized microwave fields at a frequency equal to the NV center zero-field splitting (ZFS), we utilize a hidden effective Raman coupling \cite{cerrillo2020low} to create a coherent superposition of the 
$\vert \pm1\rangle$ spin states (we denote $\left|m_S=\lambda\right\rangle = \left|\lambda\right\rangle$).
Similar approaches have been used previously to create non-invasive bio-sensors \cite{sekiguchi2016geometric}, detect high-frequency AC magnetic fields \cite{saijo2018ac}, in fluorescence thermometry \cite{toyli2013fluorescence, hodges2013timekeeping, PhysRevLett.110.130802}, and in quantum information \cite{sekiguchi2016geometric, sekiguchi2019dynamical, sekiguchi2017optical}. 
However, these relied on Ramsey and Hahn echo experiments, which have limited sensitivity, and are prone to spin population leakage \cite{sekiguchi2016geometric}.
%
%
%
%
Dynamical decoupling protocols are a well established method to detect AC fields from nearby nuclear spins \cite{glenn2018high, romach2015spectroscopy, lovchinsky2017magnetic, alvarez2011measuring, abobeih2019atomic} or artificially applied 
fields \cite{degen2017quantum}, 
due to their long coherence times.
Moreover, they are mandatory for many high sensitivity protocols, e.g. Qdyne \cite{schmitt2017submillihertz} and are used to detect J-coupling \cite{glenn2018high} or chemical shifts \cite{aslam2017nanoscale}.
They are also required for nuclear quadrupole resonance spectroscopy \cite{lovchinsky2017magnetic}.
Therefore, it is essential to implement such sequences in zero- and low-field.
However, detuning from the ZFS due to hyperfine interaction and stray magnetic fields reduces the fidelity of the pulses, leading to erroneous signals \cite{sekiguchi2019dynamical}. 
As linearly polarized microwave fields do not offer full phase control 
in the $\vert \pm1\rangle$ subsystem of the NV center's ground states, conventional dynamical decoupling protocols, e.g., the XY8 sequence \cite{gullion1990new, wang2012comparison}, are not efficient anymore \cite{sekiguchi2019dynamical}.
We overcome this problem by limiting ourselves to 
phases of $\varphi=0$ and $\varphi=\pi$, 
to 
reduce the dynamics of the three-level system to an effective two-level system due to the inherent symmetry of the former.
As a result, we construct robust pulse sequences to efficiently decouple the spin from environmental noise and create narrowband filters to sense nearby AC magnetic fields.
In addition, we use the GRAPE algorithm \cite{KHANEJA2005296} to improve performance by optimizing the amplitude and phase of pairs of pulses to be used in the sequence.
\newline
We demonstrate the first, robust dynamical decoupling sequences with linear polarized microwaves for NV centers, 
applicable in both zero- and low-field.
Our method is applicable in NV center based setups and expanded to other three-level systems, e.g., in atoms or ions \cite{degen2017quantum}.
%
%
In combination with its bio-compatibility \cite{vaijayanthimala2009biocompatibility,mohan2010vivo,mcguinness2011quantum,fang2011exocytosis,hall2013nanoscale}, small size and capability to work with nano-scale samples sizes in a broad temperature \cite{acosta2010temperature,de2020temperature} and pressure range \cite{doherty2014electronic,lesik2019magnetic}, we 
demonstrate that NV centers are a suitable alternative to conventional zero-field sensors \cite{drung1990low, trabesinger2004squid, mcdermott2002liquid, ledbetter2009optical,blanchard2020zero}. 

\section{System}
\begin{figure}
\center
\includegraphics[width = 0.48\textwidth]{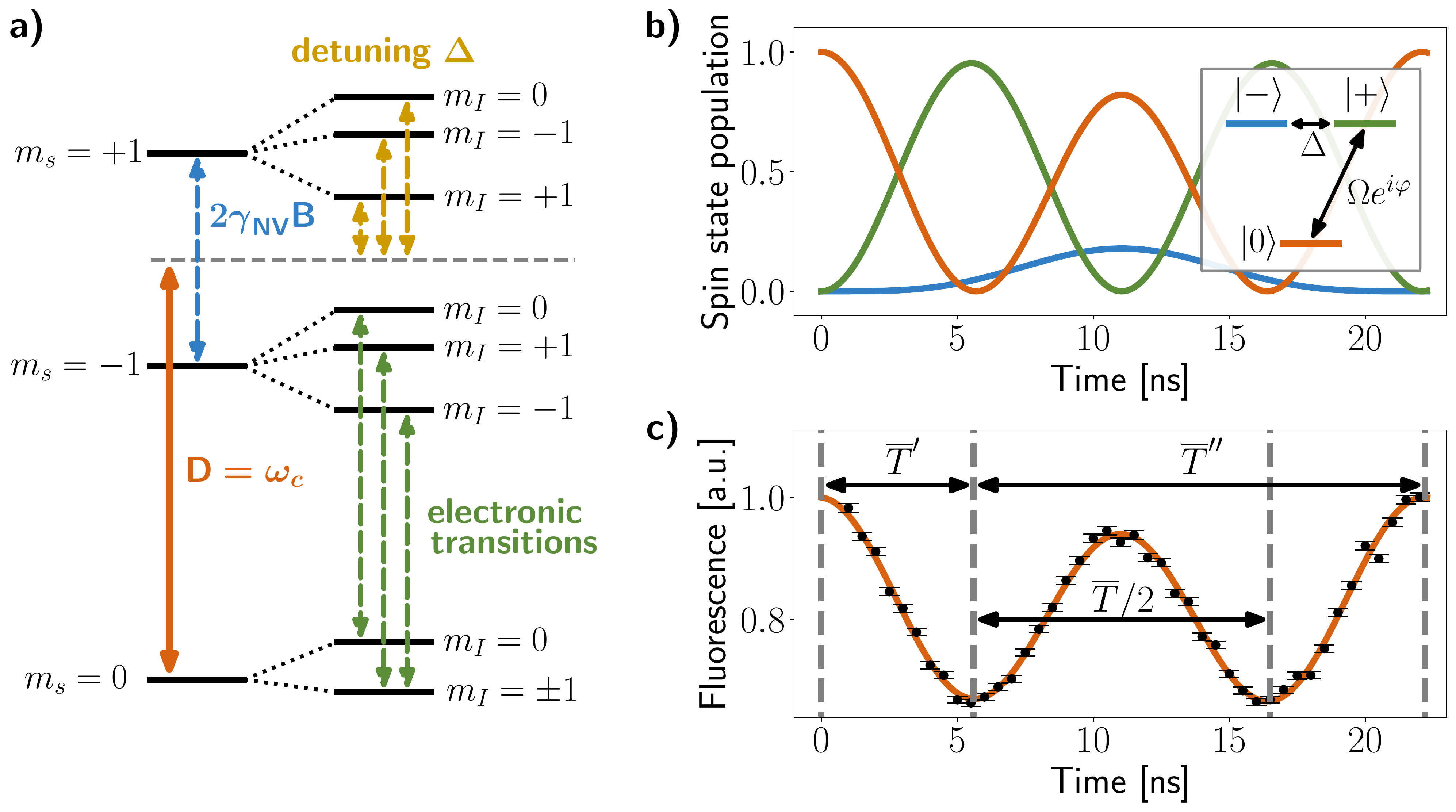}
\caption{\textbf{a)} The NV centers ground state has a zero-field splitting of $D\approx2.87~\mathrm{GHz}$. A bias magnetic field lifts the degeneracy of the $m_S=\pm1$ states by $2\gamma_{\mathrm{NV}}B$. Hyperfine coupling to the inherent ${}^{14}\mathrm{N}$ nucleus leads to an additional splitting of the spin states. We combine these detuning effects to an effective detuning $\Delta$. 
\textbf{b)} 
Microwave pulses with frequency $\omega_c = D$ lead to Rabi oscillations among the three spin states with frequencies $\overline{\Omega}=\sqrt{\Omega^2 + \Delta^2}$ and $2\overline{\Omega}$.
Depending on the ratio between detuning $\Delta$ and the applied Rabi frequency $\Omega$, part of the population will be trapped in $\vert -\rangle$.
The 
example is simulated for $\Omega=44.2~\mathrm{MHz}$ and $\Delta=9.8~\mathrm{MHz}$.
\textbf{c)} A microwave pulse with length $\overline{T}'$ flips the population from $\vert 0 \rangle$ to $\vert\phi\rangle=\left(1-\exp{\left(i\phi\right)}\right)/2\vert+\rangle-\left(1+\exp{\left(i\phi\right)}\right)/2\vert-\rangle$ \cite{cerrillo2020low}. To completely recover the population, we have to apply a microwave pulse for the time $\overline{T}''$. A pulse with length $\overline{T}/2$ acts as a conventional $\pi$-pulse in the $\vert\pm1\rangle$ subspace.}
\label{fig:level_rabi}
\end{figure}
%
The NV center is a point defect in the diamond lattice consisting of a substitutional nitrogen atom and a vacancy on the neighboring lattice side.
It's negative charge state allows 
to optically determine the electron spin state and furthermore polarize it into the $\vert0\rangle$ ground state \cite{manson2006nitrogen,doherty2013nitrogen} due to spin-selective intersystem crossing to a metastable singlet state between ground and excited state.
It possesses an $^3\mathrm{A}_2$ triplet ground state with a ZFS of $D\approx2.87~\mathrm{GHz}$, as shown in Fig. \ref{fig:level_rabi} a). 
Application of a bias magnetic field along the NV center's symmetry axis lifts the degeneracy of the $\vert\pm1\rangle$ states by $2\gamma_{\mathrm{NV}}B$.
Coupling to the inherent ${}^{14}$N nucleus and other surrounding spins causes an additional hyperfine splitting.
We combine these detuning effects and approximate them with an effective detuning $\Delta$.
If we apply a microwave $2\Omega \cos(\omega_ct+\varphi)S_x$ with a frequency $\omega_c=D$, the system's rotating frame Hamiltonian becomes
\begin{equation}
H\left(D\right)= \left(\Delta\vert-\rangle +  e^{i\varphi}\Omega\vert0\rangle\right)\langle+\vert + H.c.
\label{ep:Hrot}
\end{equation}
after the rotating wave approximation and reveals a hidden effective Raman coupling, through a change of basis to $\lbrace \vert\pm\rangle,\vert0\rangle \rbrace$ with $\vert\pm\rangle=(\vert+1\rangle\pm\vert-1\rangle)/\sqrt{2}$, sketched in figure \ref{fig:level_rabi} b).
%
%
As all three states $(\lbrace \vert\pm\rangle,\vert0\rangle \rbrace)$ are coupled for $\Delta\neq0$, continuous application of a microwave leads to an oscillation which is composed of two frequencies $\overline{\Omega}=\sqrt{\Omega^2 + \Delta^2}$ and $2\overline{\Omega}$, as shown in Fig. \ref{fig:level_rabi} b).
Due to spin flips of the nitrogen nucleus \cite{neumann2010single}, the Rabi experiment will show an additional beating on the microsecond timescale since 
$\Delta$ changes.
\\For $\Omega > \Delta$, a microwave pulse with length $\overline{T}'=\arccos(-\Delta^2/\Omega^2)/\overline{\Omega}$ will, regardless of the phase $\varphi$, flip the population from $\vert0\rangle$ to $\vert\phi\rangle=\left(1-\exp{\left(i\phi\right)}\right)/2\vert+\rangle-\left(1+\exp{\left(i\phi\right)}\right)/2\vert-\rangle$ with $\phi=\arccos\left(\frac{2\Delta^2}{\Omega^2}-1\right)$ \cite{cerrillo2020low}, as shown in Fig. \ref{fig:level_rabi} c).
Depending on the ratio $\Delta/\Omega$, part of the population will be trapped in $\vert -\rangle$.
%
In contrast to previous work \cite{sekiguchi2016geometric}, we apply a microwave pulse with length $\overline{T}''$, shown in Fig. \ref{fig:level_rabi} c). 
It allows us to fully recover the spin state population from the $\vert\pm\rangle$ manifold. 

\section{Ramsey Experiment}
\begin{figure}
\center
\includegraphics[width = 0.48\textwidth]{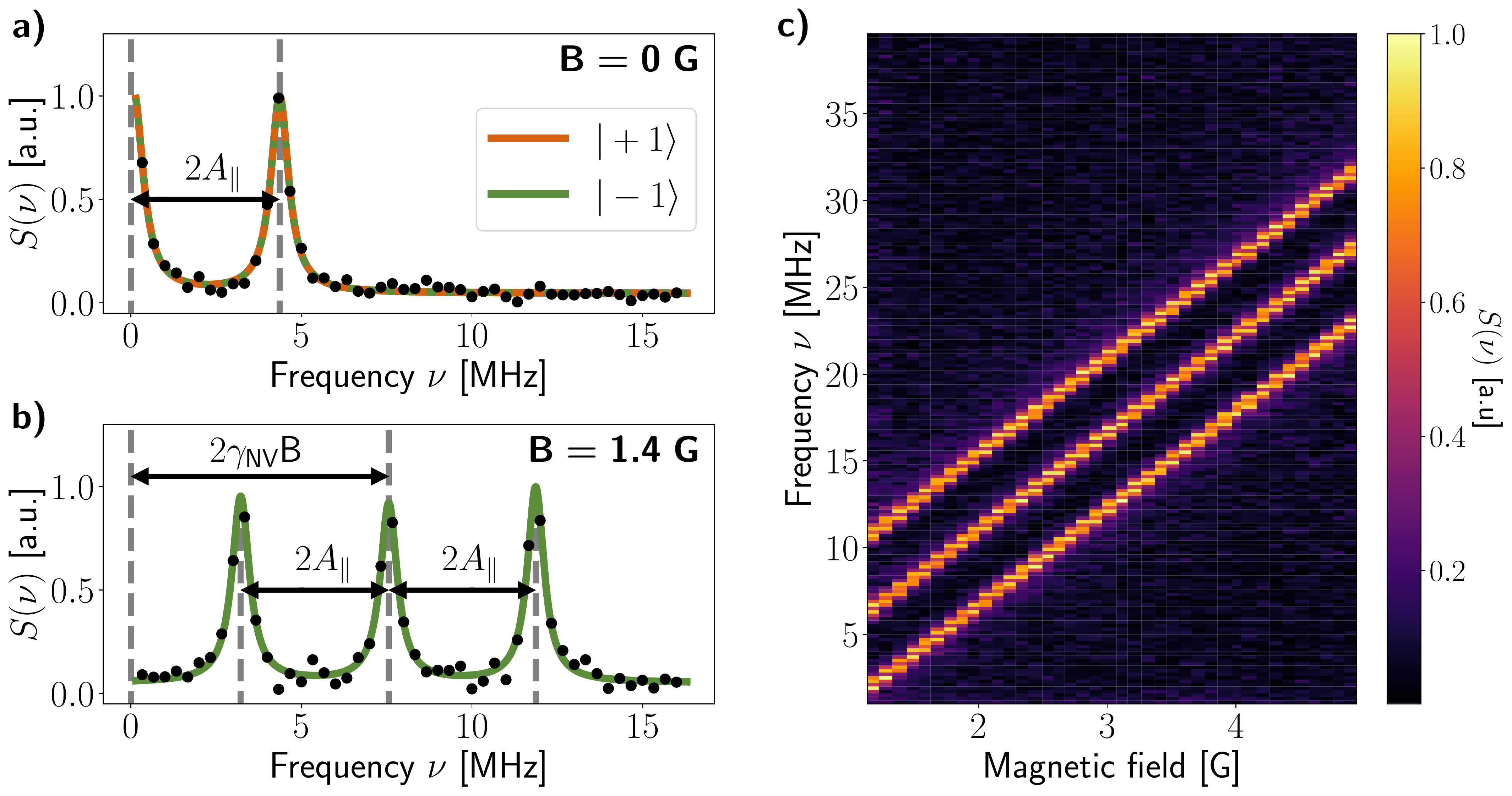}
\caption{\textbf{a)} At a magnetic field of $0~\mathrm{G}\pm 0.12~\mathrm{G}$ the $\vert\pm1\rangle$ states overlap. Each shown frequency corresponds to a hyperfine transition of the ${}^{14}\text{N}$ nucleus, which are separated by $2A_\parallel$. The uncertainty of the measurement is 
given by the line-width of the observed hyperfine frequencies. If we are at non-perfect zero-field and the exact value of the magnetic field cannot be resolved, our method is still applicable.
\textbf{b)} A bias magnetic field shifts the frequencies by $2\gamma_{\text{NV}}B$ and the spin states do not overlap anymore. Due to the double quantum transition, all observed frequencies 
are larger by a factor of two in comparison to a single quantum measurement.
\textbf{c)} Our sensing scheme is also applicable in 
low-field as long as the condition $\Omega>\Delta$ is fulfilled. The three visible, bright lines correspond to the hyperfine coupling $A_\parallel=2.166~\mathrm{MHz}~\pm~0.006~\mathrm{MHz}$ to the inherent ${}^{14}\mathrm{N}$ nucleus.}
\label{fig:ramsey}
\end{figure}
%
%
%
To demonstrate the applicability of the protocol and the full spin population recovery in zero- and low-field, we perform a Ramsey experiment and sense the inherent ${}^{14}\mathrm{N}$ nucleus \cite{haase2018controllable, sekiguchi2016geometric}.
The corresponding pulse sequence reads as $\overline{T}'-\tau-\overline{T}''$, where $\tau$ is the free evolution time.
\\All experiments are carried out in a room-temperature confocal setup with single, micron deep NV centers. 
The diamond is CVD grown (Element Six) and has a natural abundance of ${}^{13}\mathrm{C}$.
Linearly polarized microwave fields are applied through a simple wire spanned over the diamond's surface.
Both, zero- and low-field are achieved through a combination of permanent magnets, which in the latter case, are aligned with the NV center's symmetry axis. 
The zero-field is experimentally verified via a Ramsey experiment, leading to $0~\mathrm{G}\pm 0.12~\mathrm{G}$, as shown in Fig. \ref{fig:ramsey} a).
The uncertainty of the magnetic field determination is given by the linewidth of the 
hyperfine transition, as the $\vert\pm1\rangle$ states overlap.
Due to the double quantum transition, the observed frequencies are larger by a factor of two, which is best seen in Fig. \ref{fig:ramsey} b).
\\The Ramsey experiment is repeated 
up to $5~\mathrm{G}$, as shown in Fig. \ref{fig:ramsey} c). %
From the measurement we can extract the hyperfine coupling $A_\parallel=2.166~\mathrm{MHz}~\pm~0.006~\mathrm{MHz}$ to the inherent ${}^{14}\mathrm{N}$ nucleus.
In accordance to \cite{barry2020sensitivity} this leads to an estimated, shot-noise-limited sensitivity of $70~\mathrm{nT}/\sqrt{\mathrm{Hz}}~\pm~10~\mathrm{nT}/\sqrt{\mathrm{Hz}}$.
%

\section{Low-field dynamical decoupling}
\begin{figure}
\center
\includegraphics[width = 0.48\textwidth]{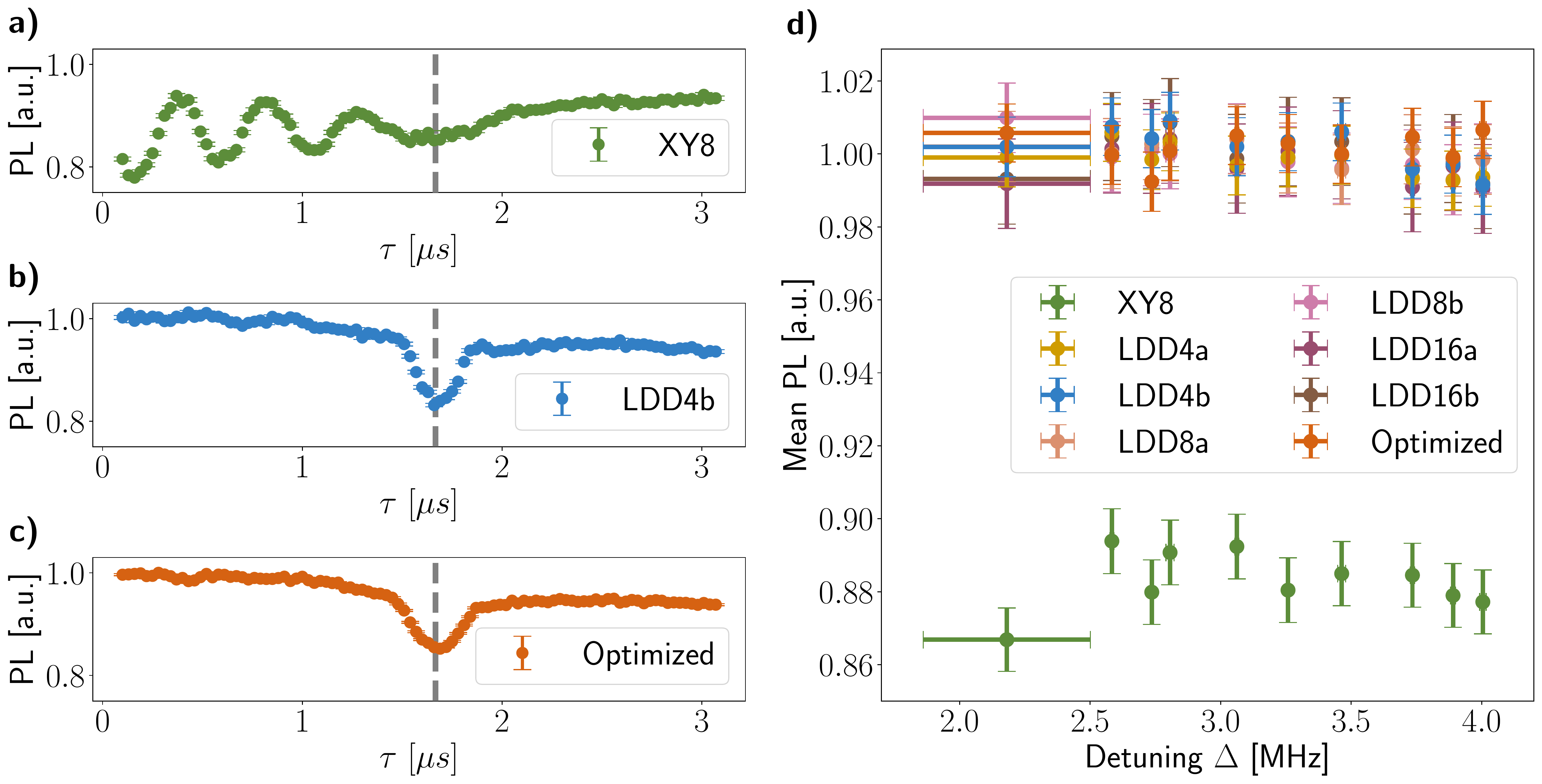} 
\caption{\textbf{a)} Pulse errors of the applied $\pi$-pulses make it impossible to detect the artificially applied 300 kHz AC signal with 
the XY8 sequence. 
The resulting erroneous signals completely overshadow the actual signal. Frequency detuning from $D$ can lead to an additional envelope \cite{sekiguchi2019dynamical}.
\textbf{b)} Due to their enhanced robustness against detuning, our optimized pulse pairs are able to overcome this problem, leading to a clearly visible signal.
\textbf{c)} Our LDD sequences are also capable of solving this problem and resolve the signal perfectly, as 
demonstrated by the LDD4b sequence.
\textbf{d)} We use the mean value of the photoluminescence 
to characterize the robustness of our pulse pairs and LDD sequences. A perfect decoupling corresponds to a value of one. All LDD sequences and the optimized pulse pairs achieve a similar robustness within the measured detuning range, matching the perfect decoupling. The XY8 sequence, however, fails to efficiently decouple the spin, clearly demonstrating the superiority of our sequences at zero- and low-field.}
\label{fig:dd}
\end{figure}

%
Dynamical decoupling protocols apply $\pi$-pulses to invert the spin state, effectively filtering out unwanted environmental noise  \cite{hahn1950spin,carr1954effects,degen2017quantum}.
%
In our double-quantum system, we use microwave $\pi$-pulses with length $\overline{T}/2=\pi/\overline{\Omega}$, as depicted in Fig. \ref{fig:level_rabi} c).
It has been observed that such $\pi$-pulses can lead to strong erroneous signals during dynamical decoupling measurements \cite{sekiguchi2019dynamical}.
An example is shown in Fig. \ref{fig:dd} a), where the applied XY8 sequence \cite{gullion1990new} is unable to resolve the artificially applied $300~\mathrm{kHz}$ AC signal. 
\\Specifically, one $\pi$-pulse generates a gate which transforms the dressed states vectors as follows:
\begin{equation}
\begin{aligned}
\vert+\rangle &\rightarrow -\vert+\rangle,\\
\vert-\rangle &\rightarrow \frac{\Omega^2-\Delta^2}{\overline{\Omega}^2}\vert-\rangle-\frac{2\Delta\Omega}{\overline{\Omega}^2}e^{i\varphi}\vert0\rangle,\\
\vert0\rangle &\rightarrow \frac{\Delta^2-\Omega^2}{\overline{\Omega}^2}\vert0\rangle-\frac{2\Delta\Omega}{\overline{\Omega}^2}e^{-i\varphi}\vert-\rangle.\\
\end{aligned}
\label{eq:state_transfer}
\end{equation}
Much like a $\pi$-pulse in a classical dynamical decoupling sequence, our $\pi$-pulse with duration $\overline{T}/2$ flips the state $\vert+\rangle$ to $-\vert+\rangle$.
However, part of the population in $\vert-\rangle$ will be shifted to $\vert0\rangle$ and vice versa unless $\Delta=0$.
A free evolution time $\tau$ in-between two $\pi$-pulses consequently creates a superposition between all three basis vectors. 
After Eq. \eqref{eq:state_transfer} the next $\pi$-pulse can then transfer even more population from the $\lbrace\vert+\rangle,\vert-\rangle\rbrace$-plane to $\vert0\rangle$, depending on the created state.
When the final $\overline{T}''$-pulse transfers the population from the $\lbrace\vert+\rangle,\vert-\rangle\rbrace$-plane back to $\vert0\rangle$, the beforehand trapped population in $\vert0\rangle$ will be flipped back to the $\lbrace\vert+\rangle,\vert-\rangle\rbrace$-plane, and we observe a loss of fluorescence. 
Depending on the detuning $\Delta$ and the chosen free evolution time $\tau$, this can result in a full loss of fluorescence, as shown in Fig. \ref{fig:dd} a).
Frequency detuning from the NV centers ZFS leads to an additional envelope \cite{sekiguchi2019dynamical}. 
In general, conventional dynamical decoupling protocols cannot overcome these problems, as the typically used phase changes do not lead to coupling between $|0\rangle$ 
and $|- \rangle$ states because of the specific rotating frame of the double quantum system (see Appendix \ref{AppendixD:TheSystem}).
Thus, we require new dynamical decoupling protocols or new pulses which offer a strong robustness against the detuning $\Delta$.
%
%
\\We introduce low-field dynamical decoupling (LDD) sequences which are engineered for dynamical decoupling at zero- and low-field.
These sequences consist of an even number of $\pi$-pulses with duration $\overline{T}/2$ whose phases $\varphi$ are used to cooperatively compensate pulse errors. 
%
%
%
A perfect $\pi$-pulse has a transition probability of 1, i.e., it inverts the population of the qubit states in the $|\pm 1\rangle$ manifold (see Appendix, sec. \ref{Sec:DD_derivation}). However, due to pulse errors, there can be an error in the transition probability, which we label $\epsilon$.
It proves useful to restrict the values of the phase to $\varphi = 0$ or $\varphi = \pi$ because we can then reduce the dynamics of the three-level system to a two-level one (see Appendix \ref{AppendixD:TheSystem}).
This simplifies the derivation significantly, allowing us to obtain analytical and numerical solutions for the two-level system and apply them directly to the zero- and low-field three-level Hamiltonian (see Appendix \ref{Sec:DD_derivation}).
\\We evaluate performance with the fidelity of the propagator in the two-level system, which characterizes the overlap between the perfect and the actual propagators \cite{genov2017prl}
$F=\frac{1}{2}\text{Tr}[(U_0^{(n)})^{\dagger}U^{(n)}]$,
where $U^{(n)}$ is the actual propagator of the pulse sequence and $U_0^{(n)}$ is its value with a perfect population inversion, i.e., when the error $\epsilon=0$. 
\\To derive the LDD phases, we perform a Taylor expansion of the error in the fidelity with respect to $\epsilon$ around $\epsilon=0$ and nullify the Taylor coefficients to the highest possible order. We obtain the simplest solution for four pulses with the LDD4a (phases: $0,0,\pi,\pi$) and LDD4b (phases: $0,\pi,\pi,0$) sequences, which correspond to the U4a and U4b sequences from \cite{genov2017prl} (see Table I). Their fidelity error is given by
$\varepsilon_{\text{LDD4}}=2\epsilon^2\sin{\left(2\widetilde{\alpha}\right)}^2$,
where $\widetilde{\alpha}$ is a phase that depends on $\Delta$, $\Omega$, and the pulse separation. This is a much smaller error in the fidelity in comparison to four pulses with zero phases $\varepsilon_{4}=8\epsilon\cos{(\widetilde{\alpha})}+O(\epsilon^2)$, allowing for cooperative error compensation. One can derive the phases for the higher order sequences in an analogous way (see Table \ref{table:coef} for the phases and Appendix \ref{Sec:DD_derivation} for derivation details).
\\
Figure \ref{fig:dd} shows the excellent performance of the LDD sequences. Due to their robustness, our optimized pulse pairs cancel the erroneous signals of Fig. \ref{fig:dd} a), and enable a precise frequency determination of the applied $300~\mathrm{kHz}$ AC signal, demonstrated in Fig. \ref{fig:dd} b). The performance of all LDD sequences and XY8 are tested for detunings up to $\Delta = 4.219~\mathrm{MHz}\pm 0.011~\mathrm{MHz}$ and demonstrate remarkable robustness, as shown in figure \ref{fig:dd} d).
We use the mean value of the photoluminescence from $\tau=0.125~\text{\textmu s}$ to $\tau=2.105~\text{\textmu s}$ with $\Delta\tau=20~\text{ns}$ step size as a measure to characterize their robustness.
A perfect decoupling corresponds to a value of one.
If a sequence is not robust to detuning, pulse errors will accumulate and greatly reduce the mean value as seen for the XY8 sequence.
All LDD sequences achieve excellent robustness, 
matching the theoretical best value, with little variation between them. 
Small deviations are caused by laser fluctuations.
This underlines the superiority of LDD sequences over classical ones. 
A further numerical comparison is found in Appendix \ref{Sec:DD_derivation}.
\begin{table}[b!]
\caption{Phases $\varphi_{k}$ of the dynamical decoupling sequences for low-field nano-NMR (LDD) with $n$ pulses (indicated by the number in the label of the sequence).
The change of sign of all phases does not change the excitation profile. All phases are defined in radians, mod($2\pi$).
}
\begin{tabular}{l l l} 
\hline 
Name~~~~~~ & Pulses~~~~~~ & Phases $\varphi_{k},~k=1\dots n$ \\ 
\hline 
LDD4a & 4 & $(0,0,1,1)\pi$ \\
LDD4b & 4 & $(0,1,1,0)\pi$ \\
LDD8a & 8 & $(0,0,1,1,0,1,1,0)\pi$ \\
LDD8b & 8 & $(0,1,1,0,0,0,1,1)\pi$ \\
LDD16a & 16 & $(0, 0, 1, 1, 0, 1, 1, 0, 1, 1, 0, 0, 1, 0, 0, 1)\pi$ \\
LDD16b & 16 & $(0, 1, 1, 0, 0, 0, 1, 1, 1, 0, 0, 1, 1, 1, 0, 0)\pi$ \\
\hline 
\end{tabular}
\label{table:coef} 
\end{table}
%
%
%
%
%
\\
We note that LDD requires a specific symmetry in the three-level Hamiltonian that allows us to reduce its dynamics to that of a two-level system (see Appendix \ref{AppendixD:TheSystem}). 
To expand the LDD applicability, we use quantum optimal control (OC) to design robust pulses that are not limited to the above symmetry, and can also be integrated within the standard LDD sequences. 
%
%
We implement GRAPE \cite{KHANEJA2005296, leung2017OptimalControl} in Julia \cite{julialanguage} using automatic differentiation \cite{innes2019differentiable, Mogensen2018}.
%
%
The optimal control pulses 
are designed to be robust against Rabi frequency and detuning errors. 
Decoupling sequences are decomposed into $\pi$-pulse pairs which, when applied consecutively, produce an identity gate. 
We task the algorithm with finding cooperative optimal control pulses \cite{BRAUN2010114} which flip the spin in the $\vert\pm 1\rangle$ manifold while not being strictly $\pi$-pulses.
%
%
The pulses are parameterized in 1 ns steps with a time-dependent amplitude and phase. The amplitude is restricted to $\Omega=20~\mathrm{MHz}$ and the phase to $\varphi=0$ or $\varphi=\pi$ as in the experiment (see Appendix \ref{Sec:DD_derivation}). Each pulse is twice the length of a standard $\pi$-pulse. 
We note that lifting the restriction on the phase did not lead to any significant improvement. 
%
%
Neglecting nuclear back action, the pulses are optimized to be robust to a detuning of $\Delta = \pm 2.16~\mathrm{MHz}$ caused by the ${}^{14}\text{N}$ nucleus, a $\pm 10~\%$ error in the Rabi frequency and a $\pm 100~\mathrm{kHz}$ shift of $D$. 
The figure of merit is taken as the overlap between the final propagator and the identity gate. 
Figure \ref{fig:dd} shows the excellent performance of the OC pulses for both sensing and in terms of robustness, similarly to LDD.

\section{Discussion}
One important application of zero- and low-field sensing are temperature measurements \cite{toyli2013fluorescence, hodges2013timekeeping}. The latter rely on the application of a $\pi$-pulse and is subject to errors in the presence of detuning $\Delta$. Our simulations and measurements show that both LDD and OC sequences achieve a very high robustness against detuning, removing all erroneous signals during the measurement in contrast to the standard pulsed dynamical decoupling (see Appendix \ref{Sec:Temp_measurements}).
\\
Many applications of NV centers also require accounting for the effect of strain 
\cite{Jamonneau2016PRB}. 
We can treat the latter as
a local static electric field interacting with the NV defect through the linear Stark effect \cite{Tamarat2006Nature}. Numerical simulations and our experiments show that the LDD sequences show improved performance in presence of strain (see Appendix \ref{Sec:Strain_effect}). 
\\
Finally, we note that LDD are not designed to compensate single quantum detuning in $D$. 
These errors can be mitigated by replacing the $\pi$-pulse in LDD by one of the OC pulses or by a robust composite pulse (see Appendix \ref{Sec:DD_derivation}, last paragraph). This can be especially useful in case of large temperature variation, leading to changes in $D$ 
to avoid leakage to $|0\rangle$. 
One can also use OC for closed-loop experimental optimization to compensate errors, which are not known or when the Hamiltonian is complex, so finding an analytical solution for time evolution of the system is not possible.  
%
%
%

%
 
%
%
\section{Conclusion}
We experimentally demonstrate the application of NV centers as precise, nano-scale sensors for zero- and low-field quantum sensing experiments.
We 
apply microwave fields at the frequency of the NV center zero-field splitting, which allows us to take advantage of a hidden effective Raman coupling to construct basic pulse gates for sensing experiments.
Detection of nearby nuclear spins via Ramsey measurements with a magnetic field up to 5 G verifies the applicability of our method in both zero- and low-field.
\\Due to pulse errors and limited phase control in the $\vert\pm1\rangle$ subsystem the standard dynamical decoupling sequences, e.g. XY8, are not efficient. Thus, we develop low-field ``LDD'' sequences by limiting pulses' phase changes to $0$ and $\pi$, which allow for robust and precise AC magnetometry at zero- and low-field.
%
Moreover, we expand their applicability and robustness via the GRAPE algorithm.
Both solutions are verified for a broad detuning range, demonstrating their superiority 
at zero- and low-field. 
As dynamical decoupling is crucial for high sensitivity experiments as Qdyne \cite{schmitt2017submillihertz} and J-coupling \cite{glenn2018high}, chemical shift \cite{aslam2017nanoscale} and NQR \cite{lovchinsky2017magnetic} measurements, our method allows expanding their working regime to zero- and low-field to investigate new systems and dynamics which were inaccessible so far.

\acknowledgments
We thank Thomas Reisser and Daniel Louzon for helpful discussions. 
G. T. G. would like to thank Bruce W. Shore for helpful advice and multiple useful discussions on coherent dynamics of multistate quantum systems. 
S. O. C. is supported by the Fundación Ramón Areces postdoctoral fellowship (XXXI edition of grants for Postgraduate Studies in Life and Matter Sciences in Foreign Universities and Research Centers), and J. P. is grateful for financial support from Grant PGC2018-097328-B-100 funded by MCIN/AEI/ 10.13039/501100011033 and, as appropriate, by “ERDF A way of making Europe”, by the “European Union”. 
J. C. acknowledges the support from Ministerio de Ciencia, Innovación y Universidades (Spain) (“Beatriz Galindo” Fellowship BEAGAL18/00081).
This project has received funding from the European Union's Horizon 2020 research and innovation program under grant agreement No 820394 (ASTERIQS) and A. M. under the Marie Sklodowska-Curie grant agreement N° 765267 (QuSCo).
This work was supported by DFG (CRC 1279 and Excellence cluster POLiS), ERC (HyperQ project), BMBF and VW Stiftung.

\appendix

\section{Sensitivity}

\begin{figure}
\center
\includegraphics[width = 0.48\textwidth]{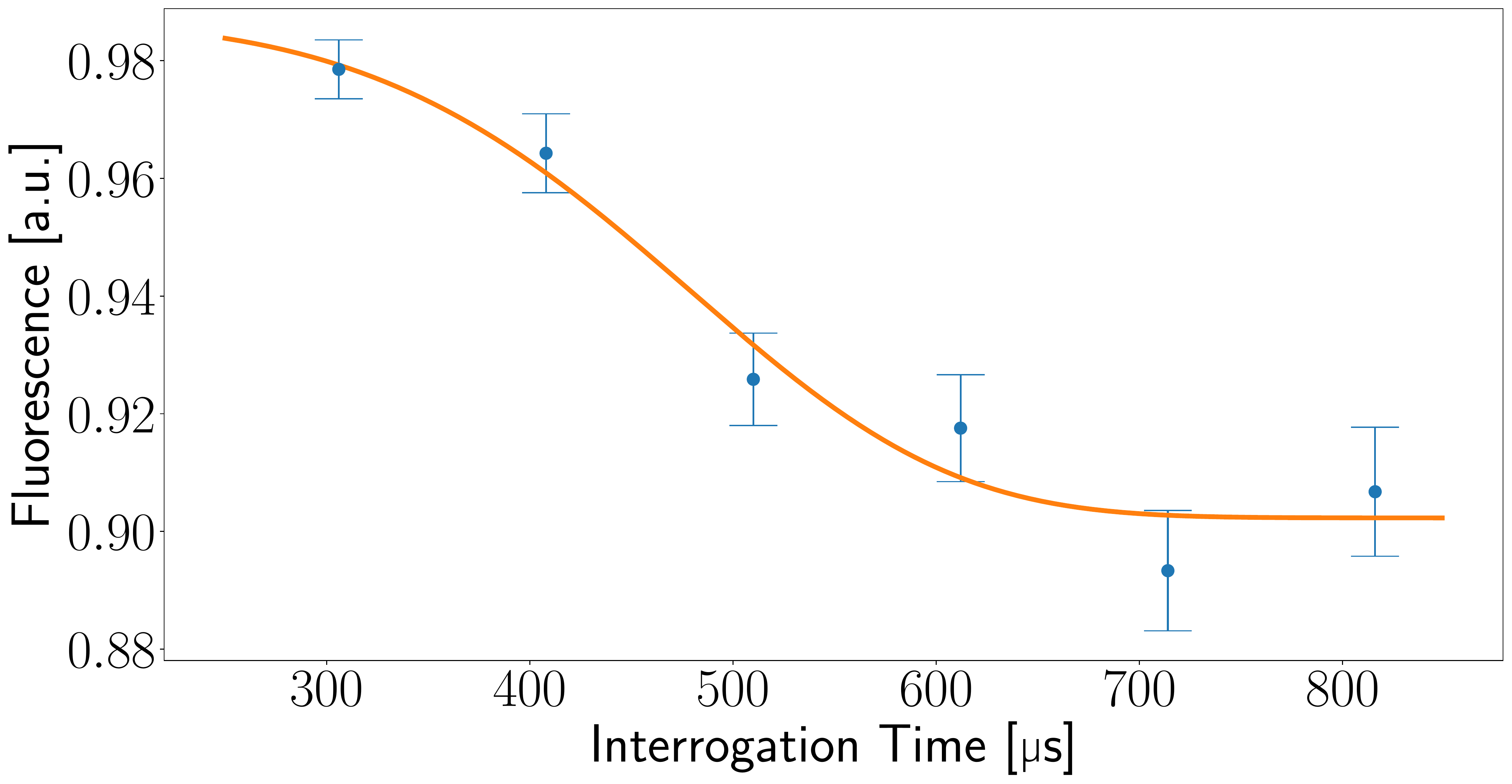}
\caption{For a frequency of 9.8 kHz, we obtain a coherence time of $500~\text{\textmu s}  \pm  40~\text{\textmu s}$ for our optimized pulse pairs. In accordance to \cite{barry2020sensitivity}, this leads to an estimated sensitivity of $\eta=8~\frac{\text{nT}}{\sqrt{\text{Hz}}}\pm3~\frac{\text{nT}}{\sqrt{\text{Hz}}}$. Due to the enhanced robustness of our optimized pulses, we can prolong the coherence time without erroneous signals. The sensitivity is highly dependent on the noise spectrum \cite{barry2020sensitivity,PhysRevB.86.045214}.}
\label{fig:sensitivity}
\end{figure}

After \cite{barry2020sensitivity}, the shot-noise limited sensitivity for a Ramsey measurement is given by
\begin{equation}
    \eta = \frac{\hbar}{\Delta m_s g_e \mu_B} \frac{1}{\sqrt{n_{avg}} C e^{-\left(\tau/T^*_2\right)^p}} \frac{\sqrt{t_I + t_r +\tau}}{\tau}.
\label{eq:sensitivity}
\end{equation}
From our measurements we obtain the following parameters:
\begin{itemize}
    \item spin quantum number difference $\Delta m_s = 2$,
    \item measurement contrast $C = 35.5~\% \pm 0.5~\%$,
    \item dephasing time $T^*_2 = 2.1~\mu\mathrm{s} \pm 0.11~\mu\mathrm{s}$,
    \item decay order $p = 2.1 \pm 0.4 $,
    \item initialization time $t_I = 22.60~\text{ns} \pm 0.28~\text{ns}$,
    \item readout time $t_R = 300~\text{ns} + 1~\mu\text{s}$,
    \item average number of photons detected per measurement $n_{avg} = 150~\frac{\text{kcounts}}{\text{s}}\cdot 300~\text{ns} \pm 10~\frac{\text{kcounts}}{\text{s}}\cdot 300~\text{ns}$.
\end{itemize}
This allows us to calculate the optimal measurement time $\tau=1.252~\mu\text{s}$, which then leads to an estimated sensitivity of $\eta=70~\frac{\text{nT}}{\sqrt{\text{Hz}}}\pm10~\frac{\text{nT}}{\sqrt{\text{Hz}}}$ for a Ramsey measurement in low-field.
\\We note that the measurement results are obtained with a $\lambda=561~\text{nm}$ laser.
\\Furthermore, we exemplary calculate the sensitivity of our optimized pulse pairs for a frequency of 9.8 kHz.
This corresponds to a $\pi$-pulse spacing of $51~\text{\textmu s}$.
By increasing the number of applied pulses we obtain a coherence time of $500~\text{\textmu s}  \pm  40~\text{\textmu s}$, as shown in figure \ref{fig:sensitivity}.
Similar to the Ramsey measurement, we can extract the measurement contrast $C = 17~\% \pm 4~\%$ and the decay order $p = 4.8 \pm 2.4$ from the measurement.
In accordance to \cite{barry2020sensitivity}, this leads to an estimated sensitivity of $\eta=8~\frac{\text{nT}}{\sqrt{\text{Hz}}}\pm3~\frac{\text{nT}}{\sqrt{\text{Hz}}}$.
We want to note that this is only an exemplary value and could be clearly optimized.
The sensitivity of dynamical decoupling sequences like our LDD sequences and optimized pulse pairs is highly depend on the environment of the investigated NV center and thus on the chosen target frequency.
The pulse distance $\tau$ needs to match the period of the target AC field and the coherence time of the dynamical decoupling sequences scales as $T_2^{\left(k\right)}=T_2k^s$ \cite{barry2020sensitivity, PhysRevB.86.045214}, with $k$ being the number of applied pulses, and $s$ is given by the noise spectrum.
%
%
%

\begin{figure*}
\includegraphics[width=\textwidth]{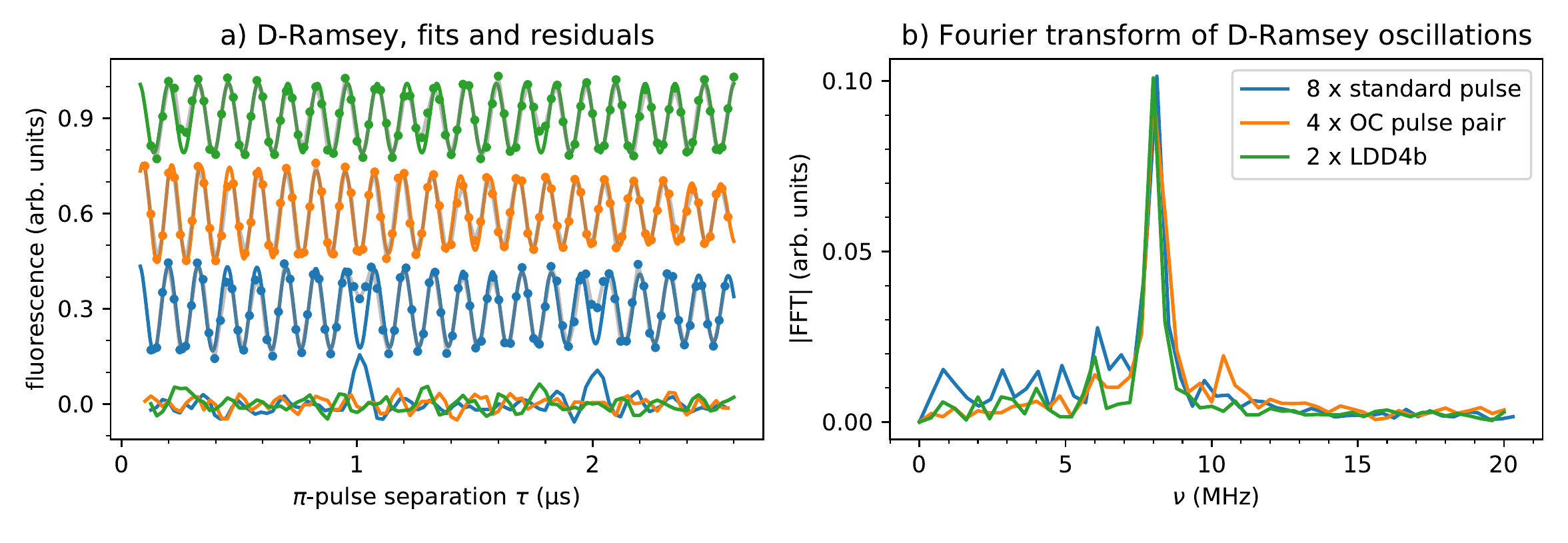}
\caption{Temperature measurements.
a) D-Ramsey oscillations with an eight $\pi$-pulse dynamical decoupling for 3 different cases (blue: standard pulses, orange: optimal control pulses, green: LDD4b pulses).
The detuning from $D$ is set to $1~\mathrm{MHz}$.
For visibility the oscillations are stacked with an offset of 0.3 from the center of the oscillation starting with 0.3 for the standard pulses.
Exponentially decaying oscillations are fit to the data. The residuals appear around 0.
b) Absolute values of the fast Fourier transform of the data from panel a.}
\label{fig:DRamsey}
\end{figure*} 

\section{Temperature measurements}\label{Sec:Temp_measurements}

In addition to magnetic field measurements, the NV center in diamond can also be used for temperature measurements.
To this end, the energy shift of the $m_S=0$ spin state from the $m_S=\pm1$ subspace is measured because the parameter $D$ is temperature sensitive ($\approx -74~\mathrm{kHz/K}$) \cite{acosta2010temperature,toyli2013fluorescence,kucsko_nanometre-scale_2013,neumann_high-precision_2013}.
When working around zero magnetic field the sensing sequence becomes fairly simple, it consists of two $\pi/4$-pulses (duration $\overline{T}'/2$) embracing a train of $\pi$-pulses (duration $\overline{T}/2$), just like in a conventional dynamical decoupling sequence \cite{hodges_timekeeping_2013,toyli2013fluorescence}.
In simple words, this sequence creates a superposition state of all three spin states.
The train of $\pi$-pulses suppresses the magnetic field induced phase acquisition of states $\ket{-1}$ and $\ket{+1}$ while state $\ket{0}$ continuously acquires phase due to its energy shift $D$.
Hence, it might be called a D-Ramsey.
Like the dynamical decoupling presented in the main paper also the temperature measurement sequence is affected by errors due to finite pulse lengths and detunings.
\\Here we compare the performance of temperature measurements using either eight standard $\pi$-pulses, four optimal control $\pi$-pulse pairs or two LDD4b repetitions.
In all cases, the microwave frequency is detuned from $D$ by about 1 MHz.
The $\pi$-pulse separation $\tau$ is swept up to $2.6~\mathrm{\mu s}$.
Figure~\ref{fig:DRamsey} a) shows the D-Ramsey oscillations vs. $\tau$ with fits of exponentially decaying oscillations to the data.
As there are eight $\pi$-pulses in each case the apparent frequency is $8~\mathrm{MHz}$ instead of $1~\mathrm{MHz}$.
Additionally, the residuals of data and fits are plotted.
For standard $\pi$-pulses these residuals reach significant values compared to the D-Ramsey oscillation amplitudes.
In temperature estimations this might lead to errors that can be mitigated by using either the LDD4b sequence or optimal control pulses.
Figure~\ref{fig:DRamsey} b) shows the absolute values of the fast Fourier transform of the data.
All measurements reveal the main frequency peak and the standard pulses produce considerable excess noise.

\section{ AC Sensing}

%
\begin{figure}[h]
\includegraphics[width = 0.48\textwidth]{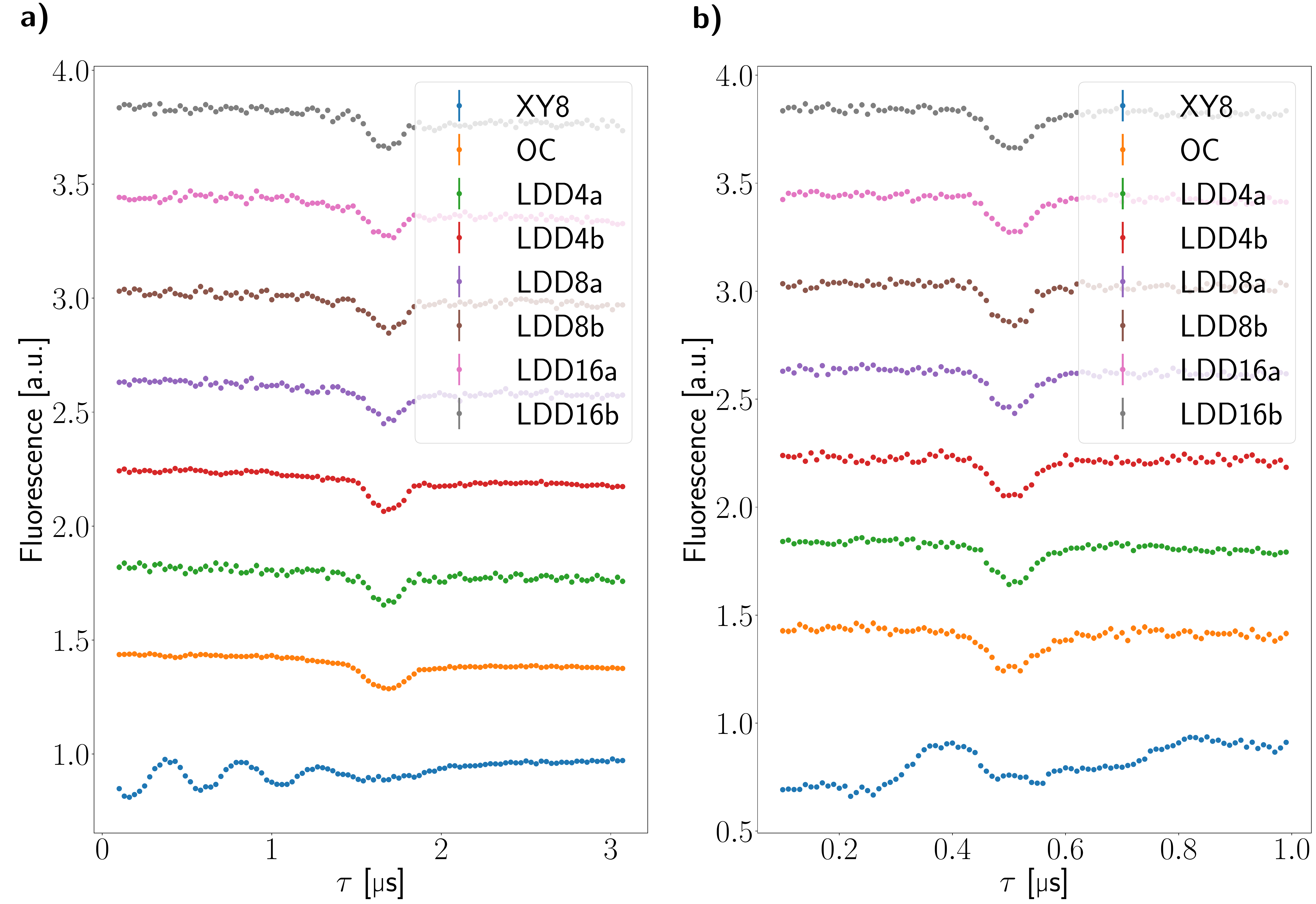}
\caption{\textbf{a)} Due to the presence of detuning, the XY8 sequence does not manage to decouple the spin efficiently, making it impossible to detect the 300 kHz AC signal. In comparison, all LDD sequences as well as our optimized pulse pairs are robust against the detuning, leading to a clearly visible signal.  \textbf{b)} The same measurements are performed for a 1 MHz AC signal. Again, our LDD sequences and optimized pulse pairs easily detect the signal, while the XY8 sequence is unable to do so.}
\label{fig:sensing}
\end{figure} 

We apply AC fields with a frequency of 300 kHz, shown in Fig. \ref{fig:sensing} a), and 1 MHz, shown in Fig. \ref{fig:sensing} b), to compare the performance of our LDD sequences and optimized pulse pairs with the XY8 sequence.
In both cases our LDD sequences and optimized pulse pairs easily detect the applied AC signal.
Their enhanced robustness against detuning lead to an efficient decoupling of the spin.
The XY8 sequence on the other hand is unable to detect either signal due to the accumulation of pulse errors, leading to strong erroneous signals.
%

\section{The System}\label{AppendixD:TheSystem}

\begin{figure}
\includegraphics[width = \columnwidth]{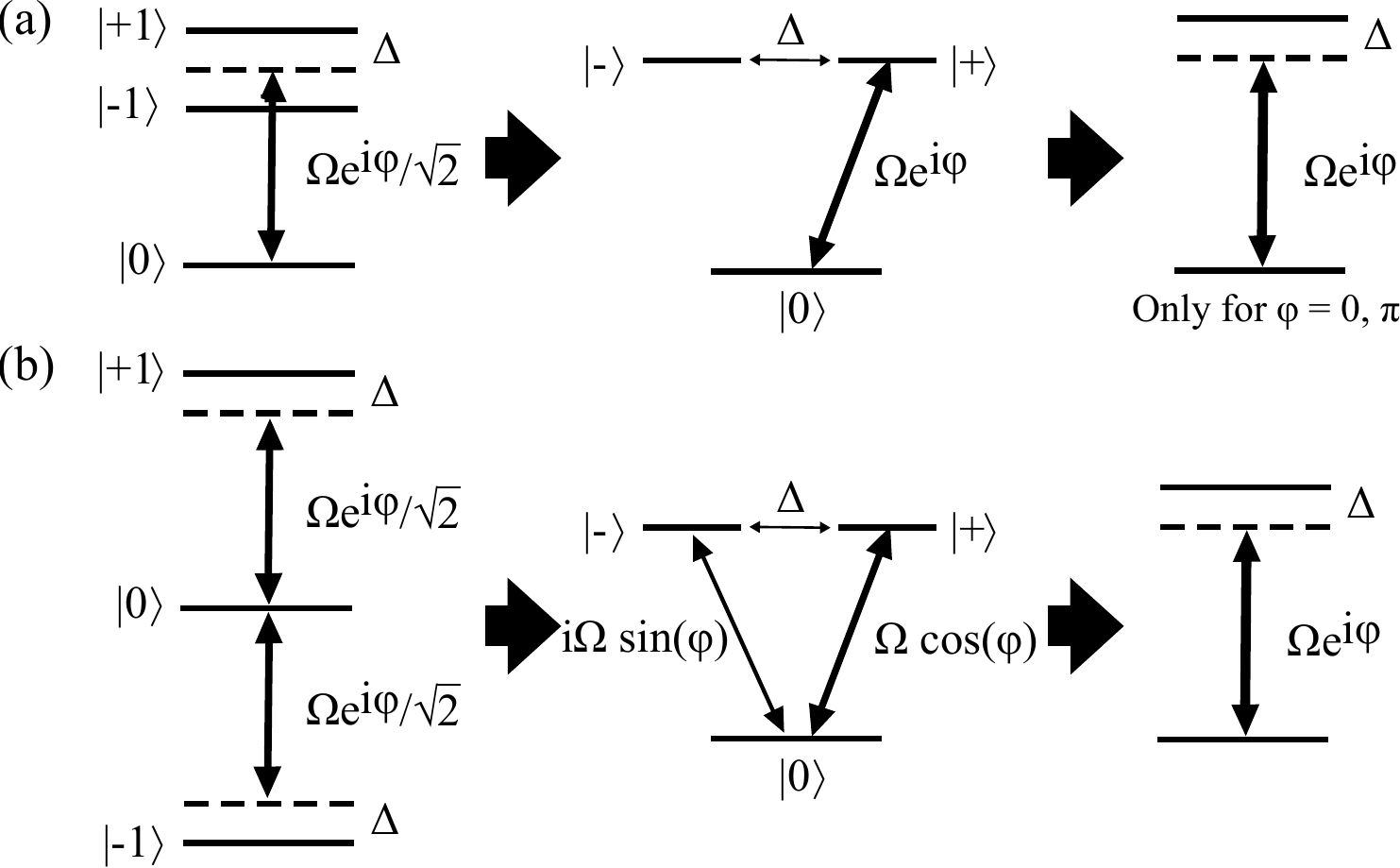}
\caption{(color online)
(a) Level scheme for low-field nano-NMR sensing, which we consider in this work and the corresponding scheme in the dressed basis with $|\pm\rangle=(|+1\rangle\pm|-1\rangle)/\sqrt{2}$. Due to the specific rotating frame, the state $|-\rangle$ is decoupled unless $\Delta\ne 0$. (b) Three-level system with SU(2) dynamic symmetry and the level scheme in the dressed basis. Due to the different definition of the rotating frame, the phase changes in $\varphi$ can lead to a coupling on the $|0\rangle\leftrightarrow|-\rangle$ transition, thus allowing for improved quantum control. The dynamics of the three-level system with SU(2) dynamic symmetry can be obtained from the evolution of the corresponding two-level system (right). This allows for direct applications of all solutions for quantum control for the two-level system, including DD sequences, to the double quantum qubit between states $|+1\rangle$and $|-1\rangle$. Such direct application is not possible in the low-field nano-NMR level scheme unless $\varphi=0$ or $\pi$, which we use for the derivation of the low field DD (LDD) sequences. Note that the $\Omega$ and $\Delta$ can in principle be time dependent.
}
\label{Fig:Level_scheme}
\end{figure}

We consider a three-level system (see Fig. \ref{Fig:Level_scheme}a) whose dynamics in the basis $\lbrace \vert +1\rangle,\vert0\rangle, \vert -1\rangle\rbrace$ is governed by the Hamiltonian 
\begin{widetext}
\begin{align}\label{H_base}
H_0&=D S_z^2+(\omega_0+\Delta) S_z+2\Omega \cos{\left(\omega_c t+\varphi\right)}S_x
\\
&=\left[ \begin{array}{ccc} D+(\omega_0+\Delta) & \Omega\sqrt{2}\cos{\left(\omega_c t+\varphi\right)} & 0 \\
\Omega\sqrt{2}\cos{\left(\omega_c t+\varphi\right)} & 0 & \Omega\sqrt{2}\cos{\left(\omega_c t+\varphi\right)} \\
0 & \Omega\sqrt{2}\cos{\left(\omega_c t+\varphi\right)} & D-(\omega_0+\Delta) \end{array}\right],\notag
\end{align}
\end{widetext}
where $S_{i},i=x, y, z$ are the spin $1$ operators, $D$ is the zero-field splitting, $\omega_0$ is the splitting between the $m=\pm 1$ states in the presence of a magnetic field, $\Delta$ is a detuning, e.g., due to a hyperfine interaction, and we apply a control field with a Rabi frequency $2\Omega$, driving frequency $\omega_c$ and phase $\varphi$.
We move to the interaction basis, defined by $H_0^{(1)}=\omega_c S_z^2$
\begin{widetext}
\begin{align}\label{H_1}
H_{\text{1}}&=U_0^{(1)}(t)^{\dagger}H_{\text{0}}
U_0^{(1)}(t)-i U_0^{(1)}(t)^{\dagger}\partial_{t}U_0^{(1)}(t)
\\
&=\left[ \begin{array}{ccc} (D-\omega_c)+\Delta+\omega_0 & \Omega\sqrt{2}\cos{\left(\omega_c t+\varphi\right)}e^{i\omega_c t} & 0 \\
\Omega\sqrt{2}\cos{\left(\omega_c t+\varphi\right)}e^{-i\omega_c t} & 0 & \Omega\sqrt{2}\cos{\left(\omega_c t+\varphi\right)}e^{-i\omega_c t} \\
0 & \Omega\sqrt{2}\cos{\left(\omega_c t+\varphi\right)}e^{i\omega_c t} & (D -\omega_c)-\Delta-\omega_0 \end{array}\right]\notag
\\
&\approx\left[ \begin{array}{ccc} \Delta & \frac{\Omega}{\sqrt{2}}e^{-i\varphi} & 0 \\
\frac{\Omega}{\sqrt{2}}e^{i\varphi} & 0 & \frac{\Omega}{\sqrt{2}}{e^{i\varphi}} \\
0 & \frac{\Omega}{\sqrt{2}}e^{-i\varphi} & -\Delta \end{array}\right]
\notag
\end{align}
\end{widetext}
where $U_0^{(1)}(t)=\exp{\left(-i \omega_c t S_z^2\right)}$ and we assumed that $\omega_0=0$ (zero-field), resonance $\omega_c=D$ and applied the rotating-wave approximation (RWA), neglecting the terms rotating at $2\omega_c t$ in the last row.

The Hamiltonian in Eq. \eqref{H_1} cannot in general be presented as a combination of the operators $S_i,i=x, y, z$ unless the phase $\varphi=0$ or $\pi$. In order to analyze the time evolution of the system, we compare it with a three-level system with a slightly different Hamiltonian, where such representation is possible for any $\varphi$ (see Fig. \ref{Fig:Level_scheme}b). Specifically,
\begin{align}\label{H_1_tilde}
\widetilde{H}_{\text{1}}&=\Delta S_z+\Omega\left(\cos{(\varphi)}S_x+\sin{(\varphi)}S_y\right)
\\
&=\left[ \begin{array}{ccc} \Delta & \frac{\Omega}{\sqrt{2}}e^{-i\varphi} & 0 \\
\frac{\Omega}{\sqrt{2}}e^{i\varphi} & 0 & \frac{\Omega}{\sqrt{2}}{\circled{$e^{-i\varphi}$}} \\
0 & \frac{\Omega}{\sqrt{2}} \circleddd{$e^{i\varphi}$} & -\Delta \end{array}\right],
\notag
\end{align}
where the elements in the dashed circles are the different ones from the Hamiltonian in Eq. \eqref{H_1}, i.e., the phase of the $|-1\rangle\leftrightarrow|0\rangle$ coupling is opposite to one in the Hamiltonian in Eq. \eqref{H_1}. This is due to the opposite direction of rotation of state $|-1\rangle$ the rotating frame of the Hamiltonian in Eq. \eqref{H_1_tilde}, which is typically $\widetilde{H}_0^{(1)}=\omega_c S_z$ in comparison to $H_0^{(1)}=\omega_c S_z^2$ for Eq. \eqref{H_1} (see Fig. \ref{Fig:Level_scheme}). The effect of this difference is not trivial and can be understood when one analyzes the system in the respective dressed basis. Specifically, the change of the phase $\varphi$ does not allow for coupling between states $|0\rangle$ and $|-\rangle$ in the dressed basis of the Hamiltonian in Eq. \eqref{H_1} unlike the case for the Hamiltonian in Eq. \eqref{H_1_tilde}, thus reducing the effect of phase changes for quantum control.

We analyze the dynamics due to the Hamiltonian in Eq. \eqref{H_1_tilde} in order to derive robust sequences of pulses for DD in the zero-field case. As it can be presented as a combination of $S_i,i=x, y, z$ operators, it is said to have SU(2) dynamic symmetry (see Fig. \ref{Fig:Level_scheme}b) and its dynamics can be characterized in terms of the dynamics of a corresponding two-state system \cite{majorana1932atomi,CookShore1979su2,hioe1987n,vitanov1997time,genov2011optimized}.
The latter is governed by the Schrodinger Eq. $i \dot{d}(t)=\widetilde{H}_{\text{2st}}d(t)$, where $d(t)=[d_1(t),d_2(t)]^{T}$ are the probability amplitudes of the two states and
\begin{equation}
\begin{aligned}
\widetilde{H}_{\text{2st}}&=\frac{\Delta}{2}\sigma_z+
\frac{\Omega}{2}\left(\cos{(\varphi)}\sigma_x+\sin{(\varphi)}\sigma_y\right)
\\&=\frac{1}{2}\left[ \begin{array}{cc} \Delta & \Omega e^{-i\varphi} \\
\Omega e^{i\varphi} & -\Delta \end{array}\right],
\end{aligned}
\label{eq:H_2st_supp}
\end{equation}
where $\sigma_i,i=x, y, z$ are the respective Pauli matrices.
The evolution of the two-state system is characterized by a propagator $\widetilde{U}_{2st}(t)$, which connects the probability amplitudes at time $t$ with ones at the initial time $t=0$: $d(t)=\widetilde{U}_{2st}(t)d(0)$ and takes the form
\begin{equation}
\widetilde{U}_{2st}=\left[ \begin{array}{cc} a & b \\ -b^{*} & a^{*} \end{array}\right],
\label{eq:Cayley-Klein_supp}
\end{equation}
%
where $a$ and $b$ are the so called Cayley-Klein parameters, which can be complex with $|a|^2+|b|^2=1$.
We note that $\Delta$ and $\Omega$ can also be time-dependent, i.e., the Cayley-Klein parameters can characterize the evolution after a Gaussian pulse, a composite pulse, or a sequence of pulses for dynamical decoupling (DD). For example, a perfect inversion of the population of the two states, e. g., by a $\pi$ pulse,  would require $a\to 0$. In addition, a perfect DD sequence, where the initial state is preserved due to refocusing of the noise, is achieved with a unity propagator, thus taking $b\to 0$.
In the case of a single pulse when they $\Omega$, $\Delta,$ and $\varphi$ are constant, the two parameters take the form:
\begin{equation}
\begin{aligned}
a &= \cos{(\overline{\Omega}t/2)}-\frac{ i\Delta}{\overline{\Omega}}\sin{\left(\overline{\Omega}t/2\right)},
\\
b &= \frac{ i\Omega e^{-i\varphi}}{\overline{\Omega}}\sin{\left(\overline{\Omega}t/2\right)},
\end{aligned}
\label{eq:Cayley-Klein-const_supp}
\end{equation}
where the effective Rabi frequency $\overline{\Omega}=\sqrt{\Omega^2+\Delta^2}$.

Next, we demonstrate the connection between the time evolution of the three-state system in Eq. \eqref{H_1_tilde} and the two-state system above, following \cite{vitanov1997time}. The former is governed by the Schrodinger equation $i \dot{c}(t)=\widetilde{H}_{\text{1}}c(t)$, where $c(t)=[c_{+1}(t),c_{0}(t),c_{-1}(t)]^{T}$ are the probability amplitudes of the system with the Hamiltonian in Eq. \eqref{H_1_tilde}.
In the case when the initial state is $c_{+1}(0)=1,~c_{0}(0)=0,~c_{-1}(0)=0$, these amplitudes can be presented in terms of the corresponding ones of the two-state system by the transformation
$c_{+1}(t)=d_1(t)^2$, $c_{0}(t)=\sqrt{2}d_1(t)d_2(t)$, $c_{-1}(t)=d_2(t)^2$, which leads to equation \eqref{eq:H_2st_supp} \cite{majorana1932atomi,CookShore1979su2,hioe1987n,vitanov1997time,genov2011optimized}. Similarly, one can obtain the
generalizations for other initial conditions and derive a propagator $\widetilde{U}_{\text{1}}(t)$, which connects the probability amplitudes at time $t$ with ones at the initial time $t=0$: $c(t)=\widetilde{U}_{\text{1}}(t)c(0)$ and characterizes the
time evolution of the three-level system for any initial state.
Given the propagator in Eq.~\eqref{eq:Cayley-Klein_supp}, which describes the evolution of the corresponding two-state system, the respective propagator for the three-state system in the basis of $\lbrace \vert +1\rangle,\vert0\rangle, \vert -1\rangle\rbrace$ is \cite{genov2011optimized}
\begin{equation}
\widetilde{U}_{\text{1}}=\left[ \begin{array}{ccc} a^{2} & \sqrt{2}ab & b^{2} \\ -\sqrt{2}ab^{*} & |a|^{2}-|b|^{2} & \sqrt{2}a^{*}b \\ b^{*2} & -\sqrt{2}a^{*}b^{*} & a^{*2} \end{array}\right].
\label{eq:Propagator3_supp}
\end{equation}
Again, we note that $\Delta$ and $\Omega$ can be time-dependent, i.e., they can characterize the evolution after a sequence of pulses for dynamical decoupling (DD). In addition, the requirement for a $\pi$ pulse in the double quantum system that inverts the population of the $\pm 1$ states is the same as for population inversion in the corresponding two-state system, namely $a\to 0$. Similarly, a perfect DD sequence that refocuses dephasing of the coherence of the double quantum qubit of states $\pm 1$ would require a unity propagator, i.e., $b\to 0$. Thus, a robust pulse or a DD sequence of pulses designed for the respective two-state system would also work for the double quantum qubit between the $\pm 1$ states the zero-field case.
Finally, we note that the zero-field Hamiltonian in Eq. \eqref{H_1} can be presented as a combination of the operators $S_i,i=x, y, z$ only for the phases $\varphi=0$ or $\pi$. Thus, we restrict our set of solutions for the DD sequences to using these two phases only and derive them for the respective two-state system. 

\section{ Derivation of the LDD sequences}\label{Sec:DD_derivation}

We derive the low-field DD (LDD) sequences by considering their effect in the absence of a sensed field. Then, unwanted errors, e.g., detuning $\Delta$ or variation in the Rabi frequency $\Omega$ cause errors in the DD sequence propagator, which leads to loss of contrast. In order to derive the LDD sequences we reparameterize the two-state system propagator in Eq. \eqref{eq:Cayley-Klein_supp} by
\begin{equation} \label{Eq:U_bare}
\widetilde{U}_{2st} = \left[\begin{array}{cc} \sqrt{\epsilon} e^{i\alpha}  & \sqrt{1-\epsilon} e^{-i\beta} \\  -\sqrt{1-\epsilon} e^{i\beta} & \sqrt{\epsilon} e^{-i\alpha}  \end{array} \right],
\end{equation}
where $p\equiv 1-\epsilon$ is the transition probability, i.e., the probability that the qubit states in the two-level system will be inverted after the interaction, $\epsilon\in[0,1]$ is the unknown error in the transition probability, $\alpha$ and $\beta$ are unknown phases.
In case of a perfect pulse, 
the transition probability becomes $p=1$ and $\epsilon=0$. However, this is often not the case, e.g., due to frequency or amplitude drifts or field inhomogeneity, 
which make $\epsilon\ne 0$. Such errors can be compensated by applying phased sequences of pulses, where the phases of the subsequent pulses are chosen to cancel the errors of the individual pulses cooperatively up to a certain order \cite{genov2014prl,genov2017prl,genov2018pra}.


%
\begin{figure*}
\includegraphics[width = 0.65\textwidth]{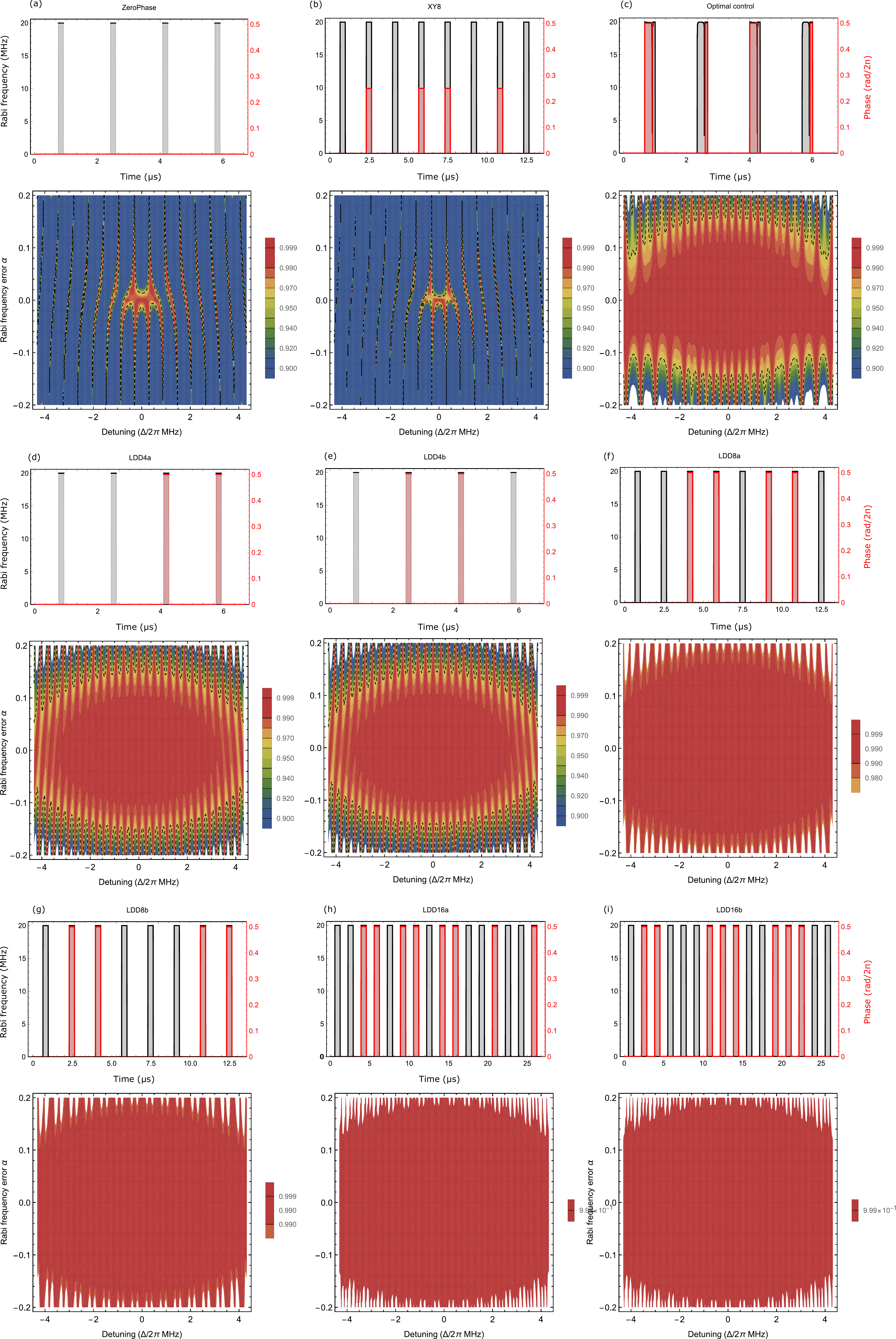}
\caption{Simulations for a Rabi frequency $\Omega=\Omega_0(1+\alpha)$, where the target magnitude of the Rabi frequency is $\Omega_0=2\pi~20$ MHz. All sequences use rectangular pulses with a duration of $25$ ns each, except for the Optimal control case, where each of the two optimized pulses is $50$ ns long. The time separation between the centers of each pulse is $\tau = 1/(2\times 300 kHz)\approx 1.67~\mu$s. The top figure shows the evolution of the pulse target Rabi frequency and phase (note that the pulse durations are longer than the actual ones for better visibility). The bottom figures show the corresponding fidelity of state $|+\rangle$ after the DD pulses for $\alpha \in [-0.2,0.2]$ and $\Delta \in 2\pi[-4.32,4.32]$ MHz for (a) Standard rectangular pulses with a zero phase , (b) the XY8 sequence, (c) DD with pulses, designed by optimal control (see text), (d) LDD4a, (e) LDD4b, and (f) LDD8a, (g) LDD8b, (h) LDD16a, (i) LDD16b. Note that the optimal control pulses were designed to achieve a high fidelity in the range of $\pm2.16~\mathrm{MHz}$ and $\pm0.1$ forvdetuning and Rabi error respectively.}
\label{fig:LDD}
\end{figure*}   
%

If the pulses are time separated, the propagator of the whole cycle $\left[\right.$free evolution for time $\tau/2-\text{pulse}-$ free evolution for time $\tau/2\left.\right]$ changes by taking $\alpha\to\widetilde{\alpha}=\alpha+\Delta\tau$. Additionally, a shift with the phase $\varphi_{k}$ at the beginning of the $k$-th pulse in a DD sequence leads to $\beta\to\beta+\varphi_{k}$ \cite{genov2014prl,genov2017prl}.
We note that we only assume the rotating wave approximation, coherent evolution, and the assumption that effect of the pulse and free evolution before and after the pulse on the qubit is the same during each pulse (except for the effect of the phase $\varphi_{k}$). Otherwise, the pulse shape can in principle be arbitrary and can also be detuned from resonance.  
Thus, the propagator of the $k$-th pulse takes the form
\begin{align}\label{Eq:U_bare_phased}
U(\varphi_k)=\left[\begin{array}{cc} \sqrt{\epsilon}e^{i\widetilde{\alpha}}  & \sqrt{1-\epsilon}e^{-i(\beta+\varphi_{k})}\\
-\sqrt{1-\epsilon}e^{i(\beta+\varphi_{k})} &\sqrt{\epsilon}e^{-i\widetilde{\alpha}} \end{array} \right].
\end{align}
Assuming coherent evolution during a sequence of $n$ pulses with different initial phases $\varphi_{k}$, the propagator of the composite sequence then becomes
\begin{equation}\label{Eq:U_phased_sequence}
U^{(n)}=U(\varphi_{n})\dots U(\varphi_{1}),
\end{equation}
and the phases $\varphi_{k}$ of the individual pulses can be used as control parameters to achieve a robust performance and can take the values $0$ or $\pi$ to correspond to the low-field system.
We can evaluate the latter by considering the fidelity \cite{genov2017prl}
\begin{equation}\label{Eq:Fid_phased_sequence}
F=\frac{1}{2}\text{Tr}\left[\left(U_0^{(n)}\right)^{\dagger}U^{(n)}\right]\equiv 1-\varepsilon_{n},
\end{equation}
where $U_0^{(n)}$ is the propagator of the respective pulse sequence when $\epsilon=0$, i.e., when the pulse performs a perfect population inversion and $\varepsilon_{n}$ is the error in the fidelity, where the label $n$ shows the number of pulses in the DD sequence. For example, the fidelity of a single pulse is given by $F=\sqrt{1-\epsilon}$. We note that this measure of fidelity does not take into account variation in the phase $\beta$, which is important when we apply an odd number of pulses. However, the latter is fully compensated when we apply an even number of pulses with perfect transition probability. Thus, we use the fidelity measure in Eq. \eqref{Eq:Fid_phased_sequence} as it usually provides a simple and sufficient measure of performance when we apply an even number of pulses. All phase shifts are assumed to be either $\phi_k=0$ or $\pi$, so that they could be used for the propagator in the zero-field case. 

First, we consider the case with two pulses in a sequence, i.e., $n=2$. The fidelity error is given by
\begin{equation}\label{Eq:Fid_2}
\varepsilon_{2}=2\epsilon\cos{\left(\widetilde{\alpha}+\frac{\varphi}{2}\right)}^2,
\end{equation}
where we assumed without loss of generality that $\varphi_1=0$. As we can see, we cannot in general reduce the error in the fidelity by choosing a particular phase $\varphi_2$ as it will work only for particular values of $\widetilde{\alpha}$ and thus only for particular detunings $\Delta$.

Second, we consider the case with four pulses in a sequence, i.e., $n=4$. In order to derive the phases we perform a Taylor expansion of the error in the fidelity with respect to $\epsilon$ around $\epsilon=0$ and nullify the first order Taylor coefficient for any $\widetilde{\alpha}$. The solution is given by the LDD4a (phases: $0,0,\pi,\pi$) and LDD4b (phases: $0,\pi,\pi,0$) sequences, which correspond to the U4a and U4b sequences from \cite{genov2017prl} (see Table I in the main text). Their fidelity error is given by
\begin{equation}\label{Eq:Fid_4}
\varepsilon_{\text{LDD4}}=2\epsilon^2\sin{\left(2\widetilde{\alpha}\right)}^2,
\end{equation}
which is much smaller error in the fidelity in comparison to the error in the fidelity for two pulses $\varepsilon_{2}\sim \epsilon$ as $\epsilon < 1$. It is also much smaller than the error for four pulses with zero phases $\varepsilon_{4}=x-x^2/8$, where $x=8\epsilon\cos{(\widetilde{\alpha})}$. We note that these solutions are not unique as the phases of these and the following LDD solutions can be shifted by an arbitrary phase and are defined mod($2\pi$). Again, we restrict ourselves only to $\varphi=0$ or $\pi$ in order for these solutions to be directly applicable for the three-level system of the zero-field Hamiltonian. 

\begin{figure*}
\includegraphics[width = 0.85\textwidth]{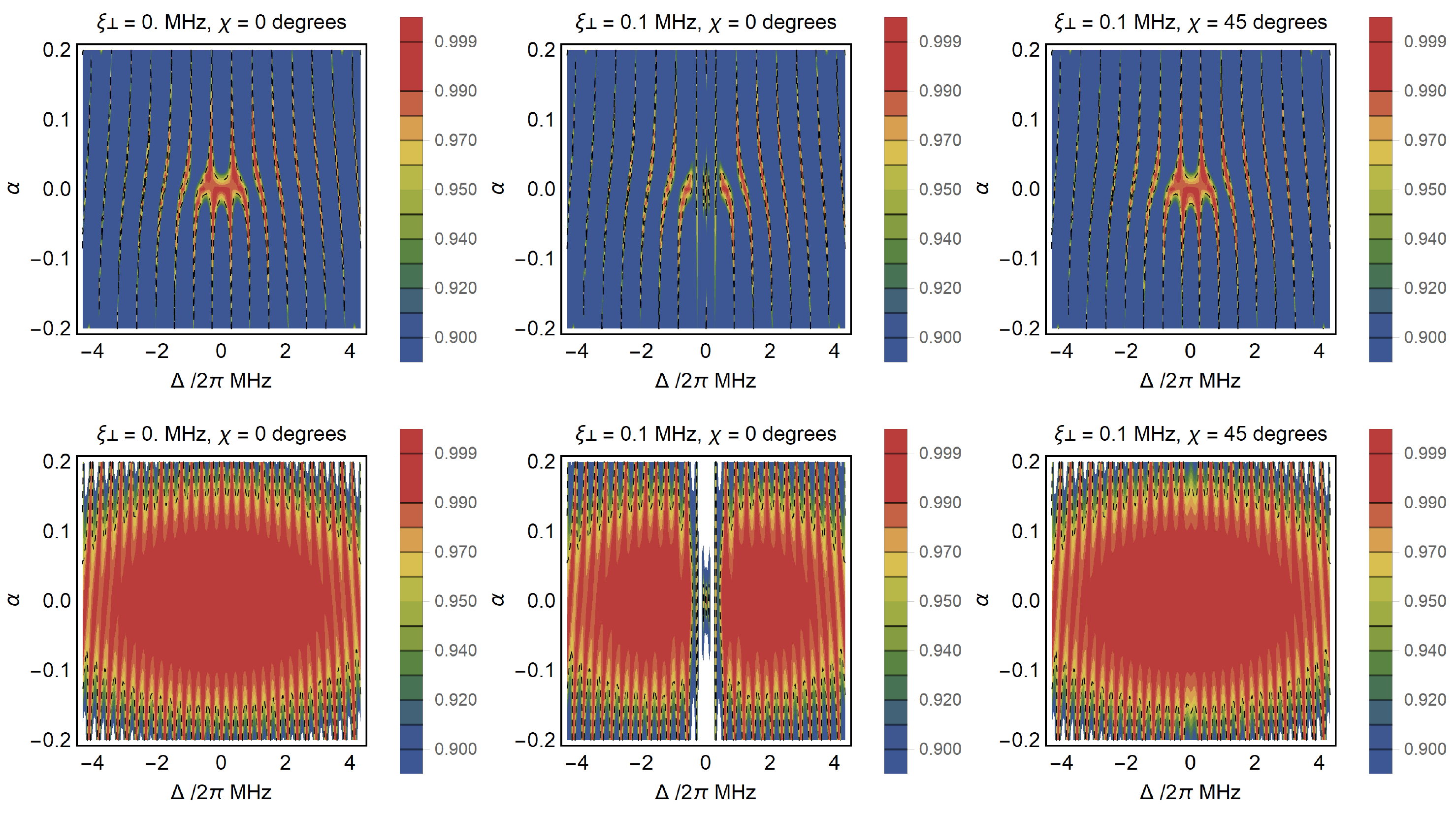}
\caption{Simulations of the fidelity of state $|+\rangle$ vs. Rabi frequency relative error $\alpha$, where $\Omega=\Omega_0(1+\alpha)$, with the target magnitude of the Rabi frequency $\Omega_0=2\pi~20$ MHz. The top figures show simulations for standard rectangular pulses with a zero phase, while the bottom figures present the corresponding simulations for LDD4a. Both sequences use rectangular pulses with a duration of $25$ ns each, where the time separation between the centers of each pulse is $\tau = 1/(2\times 300 kHz)\approx 1.67~\mu$s. The application of LDD4a leads to increased robustness. The effect of strain is slightly reduced when the driving field is perpendicular to the main component of the noise due to it, i.e., $2\chi=90$ degrees in the right figures.}
\label{fig:LDD4_strain}
\end{figure*}   

Next, we derive sequences for second order error compensation. Again, we derive the phases by performing a Taylor expansion of the error in the fidelity with respect to $\epsilon$ around $\epsilon=0$ and nullify or minimize the Taylor expansion coefficients up to the highest possible order for any $\widetilde{\alpha}$. We require $n=8$ pulses to nullify both the first and second order coefficients. There are multiple solutions and the simplest ones LDD8a and LDD8b are given in
Table I in the main text. Their phases are obtained by combining LDD4a and LDD4b one after the other and the error in the fidelity becomes
\begin{equation}\label{Eq:Fid_8}
\varepsilon_{\text{LDD8}}=8\epsilon^3\sin{\left(2\widetilde{\alpha}\right)}^2
(1-\epsilon\sin{\left(2\widetilde{\alpha}\right)}^2),
\end{equation}
which is usually much smaller than the error of the fidelity of a single LDD4 sequence or an LDD4 sequence, repeated twice, with the latter being $\varepsilon_{\text{LDD4x2}}=8\epsilon^2\sin{\left(2\widetilde{\alpha}\right)}^2
(1-\epsilon^2\sin{\left(2\widetilde{\alpha}\right)}^2)$. 

Finally, derive LDD sequences that compensate errors to the third order. We require $n=16$ pulses to nullify the first, the second and the third order coefficients of the Taylor expansion of the fidelity error. Again, there are multiple solutions and the simplest ones LDD16a and LDD16b are given in
Table I in the main text. Their phases are obtained by nesting the LDD8a and LDD8b sequences in the $(0,\pi)$ sequence and the error in the fidelity becomes
\begin{equation}\label{Eq:Fid_16}
\varepsilon_{\text{LDD16}}=8\epsilon^4\sin{\left(4\widetilde{\alpha}\right)}^2
+O(\epsilon^5),
\end{equation}
which is usually much smaller than the error of the fidelity of a single LDD8 sequence or an LDD8 sequence, repeated twice. It is also possible to derive higher order error compensating sequences in a similar way.

We note that the restriction of $\varphi=0$ or $\pi$ leads to a higher number of pulses, needed to achieve a certain order of error compensation in the two-level system in comparison to the use of arbitrary $\varphi$. For example, the UR sequences \cite{genov2014prl} use arbitrary phases and can achieve the second and third order of error compensation with $n=6$ and $n=8$ pulses, respectively. However, these are not directly applicable for the zero-field Hamiltonian.

The LDD sequences can directly be used for error compensation in the zero field three level system, defined by the Hamiltonian in Eq. \eqref{H_1} and they perform error compensation for any initial state. A numerical simulation of the fidelity for different sequences is shown in Fig. \ref{fig:LDD}, demonstrating the remarkable robustness of the sequences. We note that for the initial state $|+\rangle$ after the effective $\pi/2$ pulse in our experiment, a simple DD sequence of two pulses with $\varphi_1=0$ and $\varphi_2=\pi$ is also very efficient. However, it does not compensate errors for an arbitrary initial state. 

Finally, we note that the LDD sequences can also be combined with robust composite pulses, which can replace the simple rectangular pulses in the sequence. This could be especially useful in case of detuning errors in $D$ in order to avoid leakage to state $|0\rangle$. 
We have found in our simulations that it is particularly useful to replace the simple rectangular pulses with a composite pulse that consists of two adjacent pulses, where the first pulse has a pulse area of $3\pi/4$ with phase $0$, immediately followed by a $\pi/4$ pulse with phase $\pi$. This composite pulse acts as a robust $\pi$ pulse, uses only $\varphi=0$ or $\pi$ and has a very modest total pulse area of $2\pi$. Other longer composite pulses with alternating phases have been proposed in \cite{shaka1987symmetric}. Another alternative is to design pulses by optimal control and nest them in the LDD sequences. We note that possibly even more efficient sequences can be designed by numerical optimization but these were not necessary in our experiment.

\section{Effect of strain}\label{Sec:Strain_effect}
In many applications of zero-field sensing with NV centers it is important to take into account the effect of strain in diamond. 
Local deformations of the diamond crystal typically induce such strain, which can usually be treated as
a local static electric field interacting with the NV defect through the linear Stark effect \cite{Tamarat2006Nature}. We consider the Hamiltonian describing the NV center ground state in the presence of strain $\Sigma$, electric field $\mathbf{E}$, and magnetic field $\mathbf{B}$. Following Ref. \cite{Jamonneau2016PRB}, we also define a total electric field $\mathbf{\Pi}=\mathbf{\Sigma}+\mathbf{E}$. Thus, we can include the effect of strain on the NV center ground state Hamiltonian in Eq. \eqref{H_base} as ($\hbar=1$) \cite{Jamonneau2016PRB}
\begin{align}\label{H_strain}
H_{0,\text{s}}&=(D+d_{\parallel}\Pi_{z})S_{z}^2 +(\omega_0+\Delta) S_z+2\Omega \cos{\left(\omega_c t+\varphi\right)}S_x\notag\\
& -d_{\bot}\left(\Pi_{x}(S_{x}S_{y}+S_{y}S_{x})+\Pi_{y}(S_{x}^2-S_{y}^2)\right),
\end{align}
where $d_{\parallel}$ and $d_{\bot}$ are the longitudinal and transverse components of the electric field dipole moment 
and usually $d_{\bot}\gg d_\parallel$ \cite{Jamonneau2016PRB}. Similarly to Ref. \cite{Jamonneau2016PRB} we define $\xi_{\bot}\equiv -d_{\bot}\sqrt{\Pi_{x}^2+\Pi_{y}^2}$, $2\chi=\arctan(\Pi_{x}/\Pi_{y})$, so the Hamiltonian takes the form
\begin{widetext}
\begin{align}\label{H_base_strain}
H_{0,\text{s}}&=\left[ \begin{array}{ccc} D+d_{\parallel}\Pi_{z}+(\omega_0+\Delta) & \Omega\sqrt{2}\cos{\left(\omega_c t+\varphi\right)} & \xi_{\bot} e^{-2i \chi} \\
\Omega\sqrt{2}\cos{\left(\omega_c t+\varphi\right)} & 0 & \Omega\sqrt{2}\cos{\left(\omega_c t+\varphi\right)} \\
\xi_{\bot} e^{2i \chi} & \Omega\sqrt{2}\cos{\left(\omega_c t+\varphi\right)} & D+d_{\parallel}\Pi_{z}-(\omega_0+\Delta) \end{array}\right],
\end{align}
\end{widetext}
We move to the interaction basis, defined by $H_0^{(1)}=\omega_c S_z^2$
\begin{align}\label{H_1_strain}
H_{\text{1,s}}&=U_0^{(1)}(t)^{\dagger}H_{\text{0,s}}
U_0^{(1)}(t)-i U_0^{(1)}(t)^{\dagger}\partial_{t}U_0^{(1)}(t)\notag
\\
&\approx\left[ \begin{array}{ccc} d_{\parallel}\Pi_{z}+\Delta & \frac{\Omega}{\sqrt{2}}e^{-i\varphi} & \xi_{\bot} e^{-2i \chi} \\
\frac{\Omega}{\sqrt{2}}e^{i\varphi} & 0 & \frac{\Omega}{\sqrt{2}}{e^{i\varphi}} \\
\xi_{\bot} e^{2i \chi} & \frac{\Omega}{\sqrt{2}}e^{-i\varphi} & d_{\parallel}\Pi_{z}-\Delta \end{array}\right]
\end{align}
where $U_0^{(1)}(t)=\exp{\left(-i \omega_c t S_z^2\right)}$ and we assumed that $\omega_0=0$ (zero-field), resonance $\omega_c=D$ and applied the rotating-wave approximation (RWA), neglecting the terms rotating at $2\omega_c t$.

In order to understand the effect of strain, we rotate our basis to $\{|-1\rangle_{\chi},|0\rangle,|+1\rangle_{\chi}\}$ by the transformation $U_z(\chi)=\exp{(-i \chi S_z)}$, where the Hamiltonian in the new basis takes the form
\begin{align}\label{H_1_chi_strain}
H_{\chi,\text{s}}&=U_z(\chi)^{\dagger}H_{\text{1,s}}
U_z(\chi)\notag
\\
&=\left[ \begin{array}{ccc} d_{\parallel}\Pi_{z}+\Delta & \frac{\Omega}{\sqrt{2}}e^{-i(\varphi-\chi)} & \xi_{\bot} \\
\frac{\Omega}{\sqrt{2}}e^{i(\varphi-\chi)} & 0 & \frac{\Omega}{\sqrt{2}}{e^{i(\varphi+\chi)}} \\
\xi_{\bot} & \frac{\Omega}{\sqrt{2}}e^{-i(\varphi+\chi)} & d_{\parallel}\Pi_{z}-\Delta \end{array}\right]
\end{align}
Then, we move to the dressed basis, defined by $|\pm\rangle_{\chi}=|-1\rangle_{\chi}\pm|1\rangle_{\chi}$ and obtain
\begin{align}\label{H_d_chi_strain}
H_{\chi,\text{s,d}}&=
\left[ \begin{array}{ccc} d_{\parallel}\Pi_{z}-\xi_{\bot} & i\Omega e^{-i\varphi}\sin{(\chi)} & \Delta \\
-i\Omega e^{i\varphi}\sin{(\chi)} & 0 & \Omega e^{i\varphi}\cos{(\chi)} \\
\Delta & \Omega e^{-i\varphi}\cos{(\chi)} & d_{\parallel}\Pi_{z}+\xi_{\bot} \end{array}\right].
\end{align}
One can see that the strain effect introduces coupling between the $|0\rangle$ and $|-\rangle_\chi$ states, which is proportional to $\sin{(\chi)}$ and is not present in its absence. In addition, there is a single quantum detuning $d_{\parallel}\Pi_{z}$ between the $|0\rangle$ and each of the $|\pm\rangle_\chi$ states, and a double quantum detuning $2\xi_{\bot}$ between the $|\pm\rangle_\chi$ states. 

In order to analyze the effect of the LDD sequences, we perform a numerical simulation using the Hamiltonian in Eq. \eqref{H_1_strain}. Figure \ref{fig:LDD4_strain} shows the fidelity of state $|+\rangle$ for different detunings and coupling strength for standard pulses (top row) and the LDD4a sequence (bottom row). We consider the following strain parameters: $\xi_{\bot} = 2\pi~100$ kHz, $\chi=0$ or $\pi/2$ and  $d_{\parallel}\Pi = 2\pi~2$ kHz, which are typical for NV centers in diamond \cite{Jamonneau2016PRB}. The simulation shows that the fidelity of state $|+\rangle$ is quite sensitive to errors with standard pulses and this is also worsened slightly in the presence of strain. In contrast, LDD4a shows much improved robustness in comparison to the standard scheme and the improvement is there also in the presence of strain. The robustness is slightly improved when the driving field ($\sim S_x$) is perpendicular to the strain noise, 
i.e., $\sim S_y$ or $\chi=\pi/4$ (right figures) in comparison to the case when the noise is parallel, i.e., $\sim S_x$ or $\chi=0$ (middle figures).

\bibliographystyle{ieeetr}
\bibliography{main_prapplied.bbl}

\begin{thebibliography}{10}

\bibitem{degen2017quantum}
C.~L. Degen, F.~Reinhard, and P.~Cappellaro, ``Quantum sensing,'' {\em Reviews
  of modern physics}, vol.~89, no.~3, p.~035002, 2017.

\bibitem{balasubramanian2008nanoscale}
G.~Balasubramanian, I.~Chan, R.~Kolesov, M.~Al-Hmoud, J.~Tisler, C.~Shin,
  C.~Kim, A.~Wojcik, P.~R. Hemmer, A.~Krueger, {\em et~al.}, ``Nanoscale
  imaging magnetometry with diamond spins under ambient conditions,'' {\em
  Nature}, vol.~455, no.~7213, pp.~648--651, 2008.

\bibitem{mamin2013nanoscale}
H.~Mamin, M.~Kim, M.~Sherwood, C.~Rettner, K.~Ohno, D.~Awschalom, and D.~Rugar,
  ``Nanoscale nuclear magnetic resonance with a nitrogen-vacancy spin sensor,''
  {\em Science}, vol.~339, no.~6119, pp.~557--560, 2013.

\bibitem{lovchinsky2016nuclear}
I.~Lovchinsky, A.~Sushkov, E.~Urbach, N.~P. de~Leon, S.~Choi, K.~De~Greve,
  R.~Evans, R.~Gertner, E.~Bersin, C.~M{\"u}ller, {\em et~al.}, ``Nuclear
  magnetic resonance detection and spectroscopy of single proteins using
  quantum logic,'' {\em Science}, vol.~351, no.~6275, pp.~836--841, 2016.

\bibitem{schmitt2017submillihertz}
S.~Schmitt, T.~Gefen, F.~M. St{\"u}rner, T.~Unden, G.~Wolff, C.~M{\"u}ller,
  J.~Scheuer, B.~Naydenov, M.~Markham, S.~Pezzagna, {\em et~al.},
  ``Submillihertz magnetic spectroscopy performed with a nanoscale quantum
  sensor,'' {\em Science}, vol.~356, no.~6340, pp.~832--837, 2017.

\bibitem{aslam2017nanoscale}
N.~Aslam, M.~Pfender, P.~Neumann, R.~Reuter, A.~Zappe, F.~F. de~Oliveira,
  A.~Denisenko, H.~Sumiya, S.~Onoda, J.~Isoya, {\em et~al.}, ``Nanoscale
  nuclear magnetic resonance with chemical resolution,'' {\em Science},
  vol.~357, no.~6346, pp.~67--71, 2017.

\bibitem{pelliccione2016scanned}
M.~Pelliccione, A.~Jenkins, P.~Ovartchaiyapong, C.~Reetz, E.~Emmanouilidou,
  N.~Ni, and A.~C.~B. Jayich, ``Scanned probe imaging of nanoscale magnetism at
  cryogenic temperatures with a single-spin quantum sensor,'' {\em Nature
  nanotechnology}, vol.~11, no.~8, pp.~700--705, 2016.

\bibitem{thiel2016quantitative}
L.~Thiel, D.~Rohner, M.~Ganzhorn, P.~Appel, E.~Neu, B.~M{\"u}ller, R.~Kleiner,
  D.~Koelle, and P.~Maletinsky, ``Quantitative nanoscale vortex imaging using a
  cryogenic quantum magnetometer,'' {\em Nature nanotechnology}, vol.~11,
  no.~8, pp.~677--681, 2016.

\bibitem{jenkins2019single}
A.~Jenkins, M.~Pelliccione, G.~Yu, X.~Ma, X.~Li, K.~L. Wang, and A.~C.~B.
  Jayich, ``Single-spin sensing of domain-wall structure and dynamics in a
  thin-film skyrmion host,'' {\em Phys. Rev. Materials}, vol.~3, p.~083801, Aug
  2019.

\bibitem{glenn2018high}
D.~R. Glenn, D.~B. Bucher, J.~Lee, M.~D. Lukin, H.~Park, and R.~L. Walsworth,
  ``High-resolution magnetic resonance spectroscopy using a solid-state spin
  sensor,'' {\em Nature}, vol.~555, no.~7696, pp.~351--354, 2018.

\bibitem{joy1998relationship}
P.~Joy, P.~A. Kumar, and S.~Date, ``The relationship between field-cooled and
  zero-field-cooled susceptibilities of some ordered magnetic systems,'' {\em
  Journal of physics: condensed matter}, vol.~10, no.~48, p.~11049, 1998.

\bibitem{zheng2019zero}
H.~Zheng, J.~Xu, G.~Z. Iwata, T.~Lenz, J.~Michl, B.~Yavkin, K.~Nakamura,
  H.~Sumiya, T.~Ohshima, J.~Isoya, J.~Wrachtrup, A.~Wickenbrock, and D.~Budker,
  ``Zero-field magnetometry based on nitrogen-vacancy ensembles in diamond,''
  {\em Phys. Rev. Applied}, vol.~11, p.~064068, Jun 2019.

\bibitem{lenz2020magnetic}
T.~Lenz, A.~Wickenbrock, F.~Jelezko, G.~Balasubramanian, and D.~Budker,
  ``Magnetic sensing at zero field with a single nitrogen-vacancy center,''
  {\em Quantum Science and Technology}, 2021.

\bibitem{mrozek2015circularly}
M.~Mr{\'o}zek, J.~Mlynarczyk, D.~S. Rudnicki, and W.~Gawlik, ``Circularly
  polarized microwaves for magnetic resonance study in the ghz range:
  Application to nitrogen-vacancy in diamonds,'' {\em Applied Physics Letters},
  vol.~107, no.~1, p.~013505, 2015.

\bibitem{cerrillo2020low}
J.~Cerrillo, S.~Oviedo~Casado, and J.~Prior, ``Low field nano-nmr via
  three-level system control,'' {\em Physical Review Letters}, vol.~126,
  no.~22, p.~220402, 2021.

\bibitem{sekiguchi2016geometric}
Y.~Sekiguchi, Y.~Komura, S.~Mishima, T.~Tanaka, N.~Niikura, and H.~Kosaka,
  ``Geometric spin echo under zero field,'' {\em Nature communications},
  vol.~7, no.~1, pp.~1--6, 2016.

\bibitem{saijo2018ac}
S.~Saijo, Y.~Matsuzaki, S.~Saito, T.~Yamaguchi, I.~Hanano, H.~Watanabe,
  N.~Mizuochi, and J.~Ishi-Hayase, ``Ac magnetic field sensing using
  continuous-wave optically detected magnetic resonance of nitrogen-vacancy
  centers in diamond,'' {\em Applied Physics Letters}, vol.~113, no.~8,
  p.~082405, 2018.

\bibitem{toyli2013fluorescence}
D.~M. Toyli, F.~Charles, D.~J. Christle, V.~V. Dobrovitski, and D.~D.
  Awschalom, ``Fluorescence thermometry enhanced by the quantum coherence of
  single spins in diamond,'' {\em Proceedings of the National Academy of
  Sciences}, vol.~110, no.~21, pp.~8417--8421, 2013.

\bibitem{hodges2013timekeeping}
J.~S. Hodges, N.~Y. Yao, D.~Maclaurin, C.~Rastogi, M.~D. Lukin, and D.~Englund,
  ``Timekeeping with electron spin states in diamond,'' {\em Physical Review
  A}, vol.~87, no.~3, p.~032118, 2013.

\bibitem{PhysRevLett.110.130802}
K.~Fang, V.~M. Acosta, C.~Santori, Z.~Huang, K.~M. Itoh, H.~Watanabe,
  S.~Shikata, and R.~G. Beausoleil, ``High-sensitivity magnetometry based on
  quantum beats in diamond nitrogen-vacancy centers,'' {\em Phys. Rev. Lett.},
  vol.~110, p.~130802, Mar 2013.

\bibitem{sekiguchi2019dynamical}
Y.~Sekiguchi, Y.~Komura, and H.~Kosaka, ``Dynamical decoupling of a geometric
  qubit,'' {\em Physical Review Applied}, vol.~12, no.~5, p.~051001(R), 2019.

\bibitem{sekiguchi2017optical}
Y.~Sekiguchi, N.~Niikura, R.~Kuroiwa, H.~Kano, and H.~Kosaka, ``Optical
  holonomic single quantum gates with a geometric spin under a zero field,''
  {\em Nature photonics}, vol.~11, no.~5, pp.~309--314, 2017.

\bibitem{romach2015spectroscopy}
Y.~Romach, C.~M\"uller, T.~Unden, L.~J. Rogers, T.~Isoda, K.~M. Itoh,
  M.~Markham, A.~Stacey, J.~Meijer, S.~Pezzagna, B.~Naydenov, L.~P. McGuinness,
  N.~Bar-Gill, and F.~Jelezko, ``Spectroscopy of surface-induced noise using
  shallow spins in diamond,'' {\em Phys. Rev. Lett.}, vol.~114, p.~017601, Jan
  2015.

\bibitem{lovchinsky2017magnetic}
I.~Lovchinsky, J.~Sanchez-Yamagishi, E.~Urbach, S.~Choi, S.~Fang, T.~Andersen,
  K.~Watanabe, T.~Taniguchi, A.~Bylinskii, E.~Kaxiras, {\em et~al.}, ``Magnetic
  resonance spectroscopy of an atomically thin material using a single-spin
  qubit,'' {\em Science}, vol.~355, no.~6324, pp.~503--507, 2017.

\bibitem{alvarez2011measuring}
G.~A. {\'A}lvarez and D.~Suter, ``Measuring the spectrum of colored noise by
  dynamical decoupling,'' {\em Physical review letters}, vol.~107, no.~23,
  p.~230501, 2011.

\bibitem{abobeih2019atomic}
M.~Abobeih, J.~Randall, C.~Bradley, H.~Bartling, M.~Bakker, M.~Degen,
  M.~Markham, D.~Twitchen, and T.~Taminiau, ``Atomic-scale imaging of a
  27-nuclear-spin cluster using a quantum sensor,'' {\em Nature}, vol.~576,
  no.~7787, pp.~411--415, 2019.

\bibitem{gullion1990new}
T.~Gullion, D.~B. Baker, and M.~S. Conradi, ``New, compensated carr-purcell
  sequences,'' {\em Journal of Magnetic Resonance (1969)}, vol.~89, no.~3,
  pp.~479--484, 1990.

\bibitem{wang2012comparison}
Z.-H. Wang, G.~de~Lange, D.~Rist{\`e}, R.~Hanson, and V.~V. Dobrovitski,
  ``Comparison of dynamical decoupling protocols for a nitrogen-vacancy center
  in diamond,'' {\em Physical Review B}, vol.~85, no.~15, p.~155204, 2012.

\bibitem{KHANEJA2005296}
N.~Khaneja, T.~Reiss, C.~Kehlet, T.~Schulte-Herbrüggen, and S.~J. Glaser,
  ``Optimal control of coupled spin dynamics: design of nmr pulse sequences by
  gradient ascent algorithms,'' {\em Journal of Magnetic Resonance}, vol.~172,
  no.~2, pp.~296--305, 2005.

\bibitem{vaijayanthimala2009biocompatibility}
V.~Vaijayanthimala, Y.-K. Tzeng, H.-C. Chang, and C.-L. Li, ``The
  biocompatibility of fluorescent nanodiamonds and their mechanism of cellular
  uptake,'' {\em Nanotechnology}, vol.~20, no.~42, p.~425103, 2009.

\bibitem{mohan2010vivo}
N.~Mohan, C.-S. Chen, H.-H. Hsieh, Y.-C. Wu, and H.-C. Chang, ``In vivo imaging
  and toxicity assessments of fluorescent nanodiamonds in caenorhabditis
  elegans,'' {\em Nano letters}, vol.~10, no.~9, pp.~3692--3699, 2010.

\bibitem{mcguinness2011quantum}
L.~P. McGuinness, Y.~Yan, A.~Stacey, D.~A. Simpson, L.~T. Hall, D.~Maclaurin,
  S.~Prawer, P.~Mulvaney, J.~Wrachtrup, F.~Caruso, R.~E. Scholten, and H.~L.~C.
  L, ``Quantum measurement and orientation tracking of fluorescent nanodiamonds
  inside living cells,'' {\em Nature nanotechnology}, vol.~6, no.~6,
  pp.~358--363, 2011.

\bibitem{fang2011exocytosis}
C.-Y. Fang, V.~Vaijayanthimala, C.-A. Cheng, S.-H. Yeh, C.-F. Chang, C.-L. Li,
  and H.-C. Chang, ``The exocytosis of fluorescent nanodiamond and its use as a
  long-term cell tracker,'' {\em Small}, vol.~7, no.~23, pp.~3363--3370, 2011.

\bibitem{hall2013nanoscale}
L.~Hall, D.~Simpson, and L.~Hollenberg, ``Nanoscale sensing and imaging in
  biology using the nitrogen-vacancy center in diamond,'' {\em MRS bulletin},
  vol.~38, no.~2, pp.~162--167, 2013.

\bibitem{acosta2010temperature}
V.~M. Acosta, E.~Bauch, M.~P. Ledbetter, A.~Waxman, L.-S. Bouchard, and
  D.~Budker, ``Temperature dependence of the nitrogen-vacancy magnetic
  resonance in diamond,'' {\em Physical review letters}, vol.~104, no.~7,
  p.~070801, 2010.

\bibitem{de2020temperature}
T.~de~Guillebon, B.~Vindolet, J.-F. Roch, V.~Jacques, and L.~Rondin,
  ``Temperature dependence of the longitudinal spin relaxation time t 1 of
  single nitrogen-vacancy centers in nanodiamonds,'' {\em Physical Review B},
  vol.~102, no.~16, p.~165427, 2020.

\bibitem{doherty2014electronic}
M.~W. Doherty, V.~V. Struzhkin, D.~A. Simpson, L.~P. McGuinness, Y.~Meng,
  A.~Stacey, T.~J. Karle, R.~J. Hemley, N.~B. Manson, L.~C.~L. Hollenberg, and
  S.~Prawer, ``Electronic properties and metrology applications of the diamond
  ${\mathrm{nv}}^{\ensuremath{-}}$ center under pressure,'' {\em Phys. Rev.
  Lett.}, vol.~112, p.~047601, Jan 2014.

\bibitem{lesik2019magnetic}
M.~Lesik, T.~Plisson, L.~Toraille, J.~Renaud, F.~Occelli, M.~Schmidt,
  O.~Salord, A.~Delobbe, T.~Debuisschert, L.~Rondin, {\em et~al.}, ``Magnetic
  measurements on micrometer-sized samples under high pressure using designed
  nv centers,'' {\em Science}, vol.~366, no.~6471, pp.~1359--1362, 2019.

\bibitem{drung1990low}
D.~Drung, R.~Cantor, M.~Peters, H.~Scheer, and H.~Koch, ``Low-noise high-speed
  dc superconducting quantum interference device magnetometer with simplified
  feedback electronics,'' {\em Applied physics letters}, vol.~57, no.~4,
  pp.~406--408, 1990.

\bibitem{trabesinger2004squid}
A.~H. Trabesinger, R.~McDermott, S.~Lee, M.~M{\"u}ck, J.~Clarke, and A.~Pines,
  ``Squid-detected liquid state nmr in microtesla fields,'' {\em The Journal of
  Physical Chemistry A}, vol.~108, no.~6, pp.~957--963, 2004.

\bibitem{mcdermott2002liquid}
R.~McDermott, A.~H. Trabesinger, M.~M{\"u}ck, E.~L. Hahn, A.~Pines, and
  J.~Clarke, ``Liquid-state nmr and scalar couplings in microtesla magnetic
  fields,'' {\em Science}, vol.~295, no.~5563, pp.~2247--2249, 2002.

\bibitem{ledbetter2009optical}
M.~Ledbetter, C.~Crawford, A.~Pines, D.~Wemmer, S.~Knappe, J.~Kitching, and
  D.~Budker, ``Optical detection of nmr j-spectra at zero magnetic field,''
  {\em Journal of magnetic resonance}, vol.~199, no.~1, pp.~25--29, 2009.

\bibitem{blanchard2020zero}
J.~W. Blanchard, T.~Wu, J.~Eills, Y.~Hu, and D.~Budker, ``Zero-to
  ultralow-field nuclear magnetic resonance j-spectroscopy with commercial
  atomic magnetometers,'' {\em Journal of Magnetic Resonance}, vol.~314,
  p.~106723, 2020.

\bibitem{manson2006nitrogen}
N.~B. Manson, J.~P. Harrison, and M.~J. Sellars, ``Nitrogen-vacancy center in
  diamond: Model of the electronic structure and associated dynamics,'' {\em
  Physical Review B}, vol.~74, no.~10, p.~104303, 2006.

\bibitem{doherty2013nitrogen}
M.~W. Doherty, N.~B. Manson, P.~Delaney, F.~Jelezko, J.~Wrachtrup, and L.~C.
  Hollenberg, ``The nitrogen-vacancy colour centre in diamond,'' {\em Physics
  Reports}, vol.~528, no.~1, pp.~1--45, 2013.

\bibitem{neumann2010single}
P.~Neumann, J.~Beck, M.~Steiner, F.~Rempp, H.~Fedder, P.~R. Hemmer,
  J.~Wrachtrup, and F.~Jelezko, ``Single-shot readout of a single nuclear
  spin,'' {\em Science}, vol.~329, no.~5991, pp.~542--544, 2010.

\bibitem{haase2018controllable}
J.~F. Haase, P.~J. Vetter, T.~Unden, A.~Smirne, J.~Rosskopf, B.~Naydenov,
  A.~Stacey, F.~Jelezko, M.~B. Plenio, and S.~F. Huelga, ``Controllable
  non-markovianity for a spin qubit in diamond,'' {\em Physical review
  letters}, vol.~121, no.~6, p.~060401, 2018.

\bibitem{barry2020sensitivity}
J.~F. Barry, J.~M. Schloss, E.~Bauch, M.~J. Turner, C.~A. Hart, L.~M. Pham, and
  R.~L. Walsworth, ``Sensitivity optimization for nv-diamond magnetometry,''
  {\em Reviews of Modern Physics}, vol.~92, no.~1, p.~015004, 2020.

\bibitem{hahn1950spin}
E.~L. Hahn, ``Spin echoes,'' {\em Physical review}, vol.~80, no.~4, p.~580,
  1950.

\bibitem{carr1954effects}
H.~Y. Carr and E.~M. Purcell, ``Effects of diffusion on free precession in
  nuclear magnetic resonance experiments,'' {\em Physical review}, vol.~94,
  no.~3, p.~630, 1954.

\bibitem{genov2017prl}
G.~T. Genov, D.~Schraft, N.~V. Vitanov, and T.~Halfmann, ``Arbitrarily accurate
  pulse sequences for robust dynamical decoupling,'' {\em Phys. Rev. Lett.},
  vol.~118, p.~133202, Mar 2017.

\bibitem{leung2017OptimalControl}
N.~Leung, M.~Abdelhafez, J.~Koch, and D.~Schuster, ``Speedup for quantum
  optimal control from automatic differentiation based on graphics processing
  units,'' {\em Phys. Rev. A}, vol.~95, p.~042318, Apr 2017.

\bibitem{julialanguage}
J.~Bezanson, A.~Edelman, S.~Karpinski, and V.~B. Shah, ``Julia: A fresh
  approach to numerical computing,'' {\em SIAM Review}, vol.~59, no.~1,
  pp.~65--98, 2017.

\bibitem{innes2019differentiable}
M.~Innes, A.~Edelman, K.~Fischer, C.~Rackauckas, E.~Saba, V.~B. Shah, and
  W.~Tebbutt, ``A differentiable programming system to bridge machine learning
  and scientific computing,'' 2019.

\bibitem{Mogensen2018}
P.~K. Mogensen and A.~N. Riseth, ``Optim: A mathematical optimization package
  for julia,'' {\em Journal of Open Source Software}, vol.~3, no.~24, p.~615,
  2018.

\bibitem{BRAUN2010114}
M.~Braun and S.~J. Glaser, ``Cooperative pulses,'' {\em Journal of Magnetic
  Resonance}, vol.~207, no.~1, pp.~114--123, 2010.

\bibitem{Jamonneau2016PRB}
P.~Jamonneau, M.~Lesik, J.~P. Tetienne, I.~Alvizu, L.~Mayer, A.~Dr\'eau,
  S.~Kosen, J.-F. Roch, S.~Pezzagna, J.~Meijer, T.~Teraji, Y.~Kubo, P.~Bertet,
  J.~R. Maze, and V.~Jacques, ``Competition between electric field and magnetic
  field noise in the decoherence of a single spin in diamond,'' {\em Phys. Rev.
  B}, vol.~93, p.~024305, Jan 2016.

\bibitem{Tamarat2006Nature}
P.~Tamarat, T.~Gaebel, J.~R. Rabeau, M.~Khan, A.~D. Greentree, H.~Wilson,
  L.~C.~L. Hollenberg, S.~Prawer, P.~Hemmer, F.~Jelezko, and J.~Wrachtrup,
  ``Stark shift control of single optical centers in diamond,'' {\em Phys. Rev.
  Lett.}, vol.~97, p.~083002, Aug 2006.

\bibitem{PhysRevB.86.045214}
L.~M. Pham, N.~Bar-Gill, C.~Belthangady, D.~Le~Sage, P.~Cappellaro, M.~D.
  Lukin, A.~Yacoby, and R.~L. Walsworth, ``Enhanced solid-state multispin
  metrology using dynamical decoupling,'' {\em Phys. Rev. B}, vol.~86,
  p.~045214, Jul 2012.

\bibitem{kucsko_nanometre-scale_2013}
G.~Kucsko, P.~C. Maurer, N.~Y. Yao, M.~Kubo, H.~J. Noh, P.~K. Lo, H.~Park, and
  M.~D. Lukin, ``Nanometre-scale thermometry in a living cell,'' {\em Nature},
  vol.~500, pp.~54--58, Aug. 2013.

\bibitem{neumann_high-precision_2013}
P.~Neumann, I.~Jakobi, F.~Dolde, C.~Burk, R.~Reuter, G.~Waldherr, J.~Honert,
  T.~Wolf, A.~Brunner, J.~H. Shim, D.~Suter, H.~Sumiya, J.~Isoya, and
  J.~Wrachtrup, ``High-{Precision} {Nanoscale} {Temperature} {Sensing} {Using}
  {Single} {Defects} in {Diamond},'' {\em Nano Letters}, vol.~13,
  pp.~2738--2742, June 2013.

\bibitem{hodges_timekeeping_2013}
J.~S. Hodges, N.~Y. Yao, D.~Maclaurin, C.~Rastogi, M.~D. Lukin, and D.~Englund,
  ``Timekeeping with electron spin states in diamond,'' {\em Physical Review
  A}, vol.~87, p.~032118, Mar. 2013.

\bibitem{majorana1932atomi}
E.~Majorana, ``Atomi orientati in campo magnetico variabile,'' {\em Il Nuovo
  Cimento (1924-1942)}, vol.~9, no.~2, pp.~43--50, 1932.

\bibitem{CookShore1979su2}
R.~J. Cook and B.~W. Shore, ``Coherent dynamics of $n$-level atoms and
  molecules. iii. an analytically soluble periodic case,'' {\em Phys. Rev. A},
  vol.~20, pp.~539--544, Aug 1979.

\bibitem{hioe1987n}
F.~T. Hioe, ``N-level quantum systems with su (2) dynamic symmetry,'' {\em JOSA
  B}, vol.~4, no.~8, pp.~1327--1332, 1987.

\bibitem{vitanov1997time}
N.~Vitanov and K.-A. Suominen, ``Time-dependent control of ultracold atoms in
  magnetic traps,'' {\em Physical Review A}, vol.~56, no.~6, p.~R4377, 1997.

\bibitem{genov2011optimized}
G.~T. Genov, B.~T. Torosov, and N.~V. Vitanov, ``Optimized control of
  multistate quantum systems by composite pulse sequences,'' {\em Physical
  Review A}, vol.~84, no.~6, p.~063413, 2011.

\bibitem{genov2014prl}
G.~T. Genov, D.~Schraft, T.~Halfmann, and N.~V. Vitanov, ``Correction of
  arbitrary field errors in population inversion of quantum systems by
  universal composite pulses,'' {\em Phys. Rev. Lett.}, vol.~113, p.~043001,
  Jul 2014.

\bibitem{genov2018pra}
G.~T. Genov, D.~Schraft, and T.~Halfmann, ``Rephasing efficiency of sequences
  of phased pulses in spin-echo and light-storage experiments,'' {\em Phys.
  Rev. A}, vol.~98, p.~063836, Dec 2018.

\bibitem{shaka1987symmetric}
A.~Shaka and A.~Pines, ``Symmetric phase-alternating composite pulses,'' {\em
  Journal of Magnetic Resonance (1969)}, vol.~71, no.~3, pp.~495--503, 1987.

\end{thebibliography}

\end{document}